\documentclass[aps,twocolumn,nofootinbib,floatfix,superscriptaddress]{revtex4}
\usepackage{amsfonts}
\usepackage{amssymb}
\usepackage{amsmath}
\usepackage{graphics}
\usepackage{graphicx}
\usepackage{soul}
\usepackage{natbib}
\usepackage{bm}
\usepackage{mathtools}
\usepackage{stmaryrd}
\usepackage{epstopdf}
\usepackage{color}
\usepackage{lipsum}
\usepackage{balance}
\usepackage{enumitem}
\usepackage[colorlinks=true,citecolor=black,linkcolor=black,urlcolor=black,filecolor=black]{hyperref}

\usepackage{acronym}
\newacro{BL}{Balescu--Lenard}
\newcommand{\BL}{\ac{BL}}
\newacro{DF}{distribution function}
\newacroplural{DF}[DFs]{distribution functions}
\newcommand{\DF}{\ac{DF}}
\newcommand{\DFs}{\acp{DF}}
\newacro{HMF}{Hamiltonian Mean Field}

\newcommand{\p}{\partial}
\newcommand{\rd}{\mathrm{d}}
\newcommand{\re}{\mathrm{e}}
\newcommand{\ri}{\mathrm{i}}
\newcommand{\bw}{\mathbf{w}}
\newcommand{\bwp}{\mathbf{w}^{\prime}}
\newcommand{\Jp}{J^{\prime}}
\newcommand{\thetap}{\theta^{\prime}}
\newcommand{\mO}{\mathcal{O}}
\newcommand{\kp}{k^{\prime}}
\newcommand{\Oext}{\Omega_{\mathrm{ext}}}
\newcommand{\mB}{\mathcal{B}}
\newcommand{\Sign}{\mathrm{Sign}}
\newcommand{\half}{\tfrac{1}{2}}
\newcommand{\eps}{\epsilon}

\newcommand{\bGamma}{\boldsymbol{\Gamma}}
\newcommand{\bI}{\mathbf{I}}
\newcommand{\bM}{\mathbf{M}}
\newcommand{\bJ}{\mathbf{J}}

\newcommand{\bk}{\mathbf{k}}

\newcommand{\bO}{\mathbf{\Omega}}
\newcommand{\Ud}{U^{\mathrm{d}}}
\newcommand{\bE}{\mathbf{E}}
\newcommand{\mU}{\mathcal{U}}
\newcommand{\deltaD}{\delta_{\mathrm{D}}}
\newcommand{\mP}{\mathcal{P}}
\newcommand{\JL}{J_{L}}
\newcommand{\Uper}{U_{\mathrm{per}}}
\newcommand{\Tdyn}{T_{\mathrm{d}}}
\newcommand{\Uext}{U_{\mathrm{ext}}}
\newcommand{\Mtot}{M_{\mathrm{tot}}}

\newcommand{\Trelax}{T_{\mathrm{r}}}
\newcommand{\Tmax}{T_{\mathrm{max}}}

\newcommand{\Etot}{E_{\mathrm{tot}}}
\newcommand{\Jtot}{J_{\mathrm{tot}}}
\newcommand{\final}{\mathrm{final}}
\newcommand{\Diff}{\mathrm{Diff}}
\newcommand{\Nreal}{N_{\mathrm{real}}}
\newcommand{\KJ}{K_{J}}

\newcommand{\og}{\overline{g}}
\newcommand{\omegaR}{\omega_{\mathrm{R}}}
\newcommand{\Gcrit}{G_{\mathrm{crit}}}
\newcommand{\Nmin}{N_{\mathrm{min}}}

\newcommand{\Gmax}{G_{\mathrm{max}}}
\newcommand{\bgamma}{\boldsymbol{\gamma}}
\newcommand{\hr}{\mathrm{hr}}
\newcommand{\Min}{\mathrm{Min}}
\newcommand{\omegaRcrit}{\omega_{\mathrm{R}}^{\mathrm{crit}}}
\newcommand{\ImPart}{\mathrm{Im}}
\newcommand{\tgamma}{\widetilde{\gamma}}

\newcommand{\mL}{\mathcal{L}}
\newcommand{\intL}{\int_{\substack{ \\[0.83ex] \hphantom{-1} \\ \hspace{-1.9em} \mL}}^{\hphantom{1}} \!\!}
\newcommand{\intLb}{\int_{\substack{ \\[0.83ex] -1 \\ \hspace{-1.9em} \mL}}^{1}}
\newcommand{\intLA}{\int_{\substack{ \\[0.83ex] J_{k} + \Delta J / 2 \\ \hspace{-4.0em} \mL}}^{J_{k} + \Delta J / 2}}
\newcommand{\intLB}{\int_{\substack{ \\[0.83ex] - \Delta J / 2 \\ \hspace{-3.4em} \mL}}^{\Delta J / 2}}
\newcommand{\rest}{\mathrm{rest}}
\newcommand{\veps}{\varepsilon}

\newcommand{\cst}{\mathrm{cst.}}

\usepackage{soul} \usepackage{xcolor}

\begin{document}

\title{Kinetic blockings in long-range interacting inhomogeneous systems}

\author{Jean-Baptiste Fouvry}
\affiliation{Institut d'Astrophysique de Paris, UMR 7095, 98 bis Boulevard Arago, F-75014 Paris, France}
\author{Mathieu Roule}
\affiliation{Institut d'Astrophysique de Paris, UMR 7095, 98 bis Boulevard Arago, F-75014 Paris, France}

\begin{abstract}
Long-range interacting systems
unavoidably relax through Poisson shot noise fluctuations
generated by their finite number of particles, $N$.
When driven by two-body correlations, i.e.\ ${1/N}$ effects,
this long-term evolution is described by the inhomogeneous
Balescu--Lenard equation. Yet, in one-dimensional systems
with a monotonic frequency profile and only subject to ${ 1\!:\!1 }$ resonances,
this kinetic equation exactly vanishes:
this is a first-order \textit{full} kinetic blocking.
These systems' long-term evolution
is then driven by three-body correlations, i.e.\ ${1/N^2}$ effects.
In the limit of dynamically hot systems, this is described
by the inhomogeneous ${1/N^2}$ Landau equation.
We investigate numerically the long-term evolution of systems
for which this second kinetic equation also exactly vanishes:
this a second-order \textit{bare} kinetic blocking.
We demonstrate that these systems relax
through the ``leaking'' contributions 
of dressed three-body interactions that are neglected 
in the inhomogeneous ${1/N^2}$ Landau equation.
Finally, we argue that these never-vanishing contributions
prevent four-body correlations, i.e. ${1/N^{3}}$ effects,
from ever being the main driver of relaxation.
\end{abstract}
\maketitle

\section{Introduction}
\label{sec:Introduction}

Following an initial violent relaxation happening on 
dynamical timescales~\citep{LyndenBell1967}, long-range interacting $N$-body systems
end up on quasi-stationary states.
Force fluctuations driven by finite-$N$ shot noise
then unavoidably lead to the long-term relaxation
of these systems,
driving them ever closer to their thermodynamical equilibrium.
Fixing the system's total mass, the larger the number of particles $N$,
the slower is this relaxation.
Kinetic theory
aims at describing such long-term irreversible evolutions,
and spans a wide range of physical systems~\citep{Nicholson1992,Binney+2008,Bouchet+2012,Campa+2014}.
In the present work,
we are interested in inhomogeneous systems,
i.e.\ integrable systems with a non-trivial mean-field orbital structure~\citep{Binney+2008},
as is the case for example in galactic discs~\cite{Kormendy2013},
galactic nuclei~\cite{Alexander2017} or globular clusters~\citep{Heggie+2003}.

Limiting oneself to two-body correlations,
i.e.\ ${1/N}$ effects, the system's relaxation
is described by the inhomogeneous \BL\ equation~\citep{Heyvaerts2010,Chavanis2012}.
In the limit where collective effects are neglected,
i.e.\ dynamically hot systems,
this equation becomes the inhomogeneous Landau equation~\citep{Chavanis2013}.
When operating, the Landau equation describes a relaxation
happening on a timescale of order ${ \Tdyn N / G^{2} }$,
with $\Tdyn$ the dynamical time and $G$ the amplitude
of the pairwise interaction potential.

Yet, in inhomogeneous ${1D}$ systems
with a monotonic frequency profile
and only subject to ${ 1\!:\!1 }$ resonances,
both the Landau and \BL\ collision operators
exactly vanish,
whatever the considered mean \DF\ for the system.
This is a kinetic blocking~\citep{Eldridge+1963,Dubin2003,Bouchet+2005,Chavanis+2007,Gupta+2011,Barre+2014,Lourenco+2015}.
Such a situation cannot typically occur in higher dimensions,
where non-trivial and non-local resonances
prevent the ${1/N}$ collision operator
from exactly cancelling.
While undergoing a kinetic blocking,
systems can only evolve under the weaker contributions
of three-body correlations, i.e.\ through ${1/N^{2}}$ effects.
In the limit where collective effects can be neglected
(i.e.\ the dynamically hot limit),
\cite{Fouvry2022} derived an inhomogeneous self-consistent closed
kinetic equation describing this relaxation
which occurs on a timescale
of order ${ \Tdyn N^{2} / G^{4} }$.
We refer to this equation as the inhomogeneous ${1/N^{2}}$ Landau equation.

Interestingly, \cite{Fouvry2022} pointed out the existence
of a class of interaction potentials for which
the inhomogeneous ${1/N^{2}}$ Landau equation also exactly vanishes --
whatever the considered mean \DF\@.
We call this a second-order \textit{bare} kinetic blocking.
In the present paper, we are interested in the long-term evolution
of such peculiar systems.
Placing ourselves in the dynamically hot limit,
we show that their relaxation occurs in fact on a (slower) timescale
of order ${ \Tdyn N^{2}/G^{6} }$.
We detail how such a scaling stems from a ``leakage''
of collective effects.
We also argue that even when undergoing
a second-order bare kinetic blocking,
these systems' relaxation is still dominated by ${1/N^{2}}$ effects
and not by four-body correlations, i.e.\ ${1/N^{3}}$ effects.
Consequently, we claim that the (still unknown) inhomogeneous ${1/N^{2}}$
\BL\@ equation cannot vanish and that its contribution
always dominates over contributions from four-body correlations.

The paper is organised as follows.
In Sec.~\ref{sec:System}, we detail our system.
In Sec.~\ref{sec:KineticBlocking}, we recall the
kinetic equations at play and explain the various 
kinetic blockings.
In Sec.~\ref{sec:NumericalMeasurements}, we investigate numerically the long-term relaxation
of the considered system.
We conclude in Sec.~\ref{sec:Conclusion}.
Throughout the main text, technical details are kept to a
minimum and deferred to Appendices or to relevant references.

\section{The system}
\label{sec:System}

Let us consider a set of $N$ particles
of individual mass ${ \mu \!=\! \Mtot / N }$
with $\Mtot$ the system's (fixed) total mass.
We denote the ${1D}$ canonical (specific)
phase space coordinates ${ \bw \!=\! (\theta, J) }$,
with $\theta$ the ${2\pi}$-periodic angle
and $J$ the associated action~\citep{Binney+2008}.
We consider a total specific Hamiltonian given by
\begin{equation}
H = \sum_{i = 1}^{N} \Uext (\bw_{i}) + \sum_{i < j}^{N} \mu \, U (\bw_{i}, \bw_{j}) , 
\label{def_Htot}
\end{equation}
with ${ \Uext (\bw) }$ some given external potential
and ${ U (\bw , \bwp) }$ some pairwise interaction potential,
whose typical amplitude is denoted $G$.
We assume the symmetry 
${ U (\bw , \bwp) \!=\! U (|\theta \!-\! \thetap| , \{ J , \Jp \}) }$,
which guarantees usual conservation laws
and ensures that only ${1\!:\!1}$ resonances 
can drive relaxation.

The system's statistics is described via
the \DF\@, ${ F \!=\! F (\bw,t) }$,
normalised so that ${ \!\int\! \rd \bw F \!=\! \Mtot }$.
We assume quasi-stationarity,
i.e.\ ${ F \!=\! F (J , t)}$ only depends on the action
and the time $t$.
Similarly, the mean Hamiltonian,
${ H_{0} (\bw) \!=\! \Uext (\bw) \!+\! \!\int\! \rd \bwp U (\bw,\bwp) F(\bwp) }$,
satisfies ${ H_{0} \!=\! H_{0} (J) }$.
From it, we can define the mean orbital frequencies
${ \Omega(J) \!=\! \rd H_{0} / \rd J }$.

\begin{figure}[htbp!]
    \begin{center}
    \includegraphics[width=0.45\textwidth]{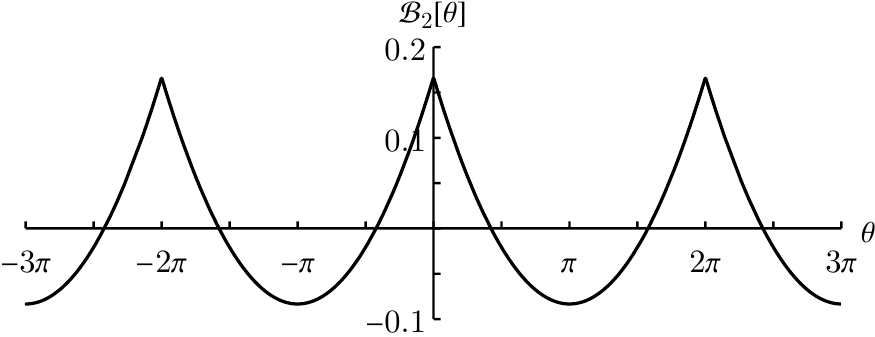}
    \caption{Angular dependence, ${ \mB_{2} (\theta) }$,
from the interaction potential of Eq.~\eqref{def_U_B2}.
    \label{fig:B2}}
    \end{center}
\end{figure}

For the rest of this work,
we will focus on one particular interaction potential,
namely eq.~{(D6)} of~\cite{Fouvry2022}.
This interaction potential allows one 
to design kinetically blocked systems 
as we will detail in Sec.~\ref{sec:KineticBlocking}.
It reads
\begin{equation}
U (\bw , \bwp)  = G \, (J \!-\! \Jp)^{2} \, \mB_{2} [ \theta \!-\! \thetap] ,
\label{def_U_B2}
\end{equation}
with ${ \mB_{2} (\theta) \!=\! B_{2} [ \tfrac{1}{2\pi} w_{2\pi} (\theta) ] }$,
${ B_{2} (x) \!=\! x^{2} \!-\! x \!+\! \tfrac{1}{6} }$
the second Bernoulli polynomial,
and the angle ``wrapping function"
\begin{equation}
0 \!\leq\! w_{2\pi} (\theta) < 2 \pi ;
\quad
w_{2\pi} (\theta) \equiv \theta \; [2 \pi] .
\label{def_wrapping}
\end{equation}
The function ${ \mB_{2} (\theta) }$ is illustrated in Fig.~\ref{fig:B2}.
We note that ${ \!\int\! \rd \theta \mB_{2}[\theta] \!=\! 0 }$,
so that ${ F (J) }$ never generates any mean potential.
Mean field quantities such as the frequency profile 
are therefore fully determined by the 
external potential ${ \Uext (\bw)}$.

In this paper, we investigate
the long-term evolution of systems driven by Eq.~\eqref{def_U_B2}
in the dynamically hot limit.
In that context, it corresponds to the limit ${G\!\ll\!\Gcrit}$
(abusively denoted ${ G \!\to\! 0 }$)
where $\Gcrit$ stands for the critical value of $G$
above which the system becomes linearly unstable
(see Appendix~\ref{app:LinearResponse}).

In order to highlight various regimes of relaxation,
we explore three different external potentials,
i.e.\ three different frequency profiles.
More precisely, fixing the prefactors to unity,
we consider
\begin{subequations}
\begin{align}
{} & \textbf{Frequency profiles}
\nonumber
\\
{} & \quad\quad\quad \text{(1):} \;\;\; \Omega (J) = |J| ;
\label{freq_1}
\\
{} & \quad\quad\quad \text{(2):} \;\;\; \Omega (J) = J \, |J| ;
\label{freq_2}
\\
{} & \quad\quad\quad \text{(3):} \;\;\; \Omega (J) = J .
\label{freq_3}
\end{align}
\label{freq_profile}\end{subequations}
And for each case, we will consider the same initial \DF\@,
${F(J)\!\propto\!\exp(-J^4)}$,
which does not correspond to the thermal equilibrium
of any of these profiles.

\section{Kinetic blockings}
\label{sec:KineticBlocking}

Given some interaction and external potentials,
kinetic  theory aims at predicting ${ \p F (J , t) / \p t }$,
i.e.\ the rate of orbital redistribution,
in the statistical limit ${N\!\gg\!1}$.
Let us now sketch the equations describing
the evolution of the previous systems
at successive orders in ${1/N}$,
highlighting in particular how kinetic blockings may occur.

\subsection{First-order kinetic equation}
\label{sec:1stKinetic}

Accounting only for two-body correlations,
assuming linear stability,
and neglecting collective effects,
the system's relaxation is described by the inhomogeneous Landau
equation~\citep{Chavanis2013}.
Limiting ourselves to ${1 \!:\! 1}$ resonances, it reads
\begin{align}
\frac{\p F(J)}{\p t} \propto \mu \frac{\p }{\p J} \bigg[ \!\sum_{k_1} k_1 {} &
\!\! \int \!\! \rd J_{1} \, \big| U_{k_1} (\bJ ) \big|^{2} \,  
\label{firstorder_Landau_eq}
\\
\times {} & \deltaD\big( \bk \!\cdot\! \bO\big) \, 
\bk\!\cdot\!\frac{\p }{\p \bJ}F_2(\bJ) \bigg] ,
\nonumber
\end{align}
where the time dependence was omitted for clarity
(see Appendix~\ref{app:1NKE} for the full expression).
We recall that ${ \mu \!=\! \Mtot / N }$,
i.e.\ this relaxation is driven by ${1/N}$ effects.
In Eq.~\eqref{firstorder_Landau_eq}, 
we shortened the notations
using the $2$-vectors ${ \bJ \!=\! (J , J_{1}) }$,
${ \bO \!=\! (\Omega[J], \Omega[J_{1}]) }$ and
${ \bk \!=\! (k_{1}, - k_{1}) }$, and 
${ F_2(\bJ) \!=\! F(J)F(J_1)}$.
This equation also involves the bare coupling coefficients,
${ U_{k_1} (\bJ) }$,
namely the Fourier transform in angles 
of the pairwise interaction potential (see Appendix~\ref{app:1NKE}).
When taking collective effects into account,
Eq.~\eqref{firstorder_Landau_eq} becomes the inhomogeneous 
\BL\ equation~\citep{Heyvaerts2010,Chavanis2012}.
It follows from Eq.~\eqref{firstorder_Landau_eq}
with the substitution
${ |U_{k_1} (\bJ)|^2 \!\to\! |\Ud_{k_1}[F](\bJ)|^2}$,
with the dressed coupling coefficients ${\Ud_{k_1}[F](\bJ)}$ 
detailed in Appendix~\ref{app:1NKE}.

For a system with a monotonic
frequency profile, ${ J \!\mapsto\! \Omega (J) }$,
the resonance condition ${ \deltaD\big(\bk\!\cdot\!\bO\big) }$
in Eq.~\eqref{firstorder_Landau_eq} imposes ${ J_{1} \!=\! J }$,
so that the cross term, ${ \bk\!\cdot\!\p F_2/\p\bJ }$,
exactly vanishes.
Ultimately, this leads to ${ \p F (J) / \p t \!=\! 0 }$,
i.e.\ the kinetic equation predicts no relaxation.
Importantly, assuming that the frequency profile is monotonic,
we stress that this cancellation
holds (i) whatever the considered interaction potential, ${ U (\bw , \bwp) }$;
(ii) whatever the considered (stable) \DF\@, ${ F (J) }$;
(iii) and for both the Landau and \BL\ equations,
i.e.\ independently of whether collective effects
are or are not accounted for.
This is a first-order \textit{full} kinetic blocking:
such systems cannot relax via two-body correlations
(${1/N}$ effects).
In that case,
the relaxation is greatly delayed
and can only occur through three-body correlations (${ 1/N^{2} }$ effects).

\subsection{Second-order kinetic equation}
\label{sec:2ndKinetic}

Placing themselves within this regime
and neglecting collective effects,
\cite{Fouvry2022} derived a closed kinetic equation
describing relaxation driven by ${1/N^{2}}$ effects.
This inhomogeneous ${1/N^{2}}$ Landau equation
is of the form
\begin{align}
    \frac{\p F(J)}{\p t} \propto \mu^{2} \frac{\p }{\p J} 
    \bigg[ \sum_{k_{1},k_{2}} {} & (k_1 \!+\!k_2) 
    \!\! \int \!\! \rd J_{1} \rd J_{2} \, |\Lambda_{k_{1} k_{2}} (\bJ)|^2 
    \label{secondorder_Landau_eq}
    \\
    {} & \times \deltaD\big(\bk \!\cdot\! \bO\big) \, 
    \bk\!\cdot\!\frac{\p }{\p \bJ}F_3(\bJ)  \bigg]  ,
\nonumber
\end{align}
and we refer to Appendix~\ref{app:1N2Landau} for the full expression of
the equation and the coupling coefficients, ${ | \Lambda_{k_{1}k_{2}}(\bJ) |^{2}}$.
In Eq.~\eqref{secondorder_Landau_eq},
notations are shortened using here the 3-vectors
${ \bJ \!=\! (J , J_{1} , J_{2}) }$,
${ \bO \!=\! (\Omega[J], \Omega[J_{1}], \Omega[J_{2}]) }$ and
${ \bk \!=\! (k_{1} \!+\! k_{2} , - k_{1} , - k_{2}) }$,
and ${ F_3(\bJ) \!=\! F(J)F(J_1)F(J_2)}$.
Since collective effects have been neglected,
it is crucial to note that the coupling coefficients,
${ |\Lambda_{k_{1}k_{2}} (\bJ)|^{2} }$,
only depend on the pairwise interaction potential:
they do \textit{not} involve the system's \DF\@, ${ F (J) }$.

The generalisation of Eq.~\eqref{secondorder_Landau_eq}
to account for collective effects,
i.e.\ the inhomogeneous ${1/N^{2}}$ \BL\ equation,
is currently unknown.
In particular, at order ${1/N^2}$,
one may expect for collective effects
to be more involved than a simple dressing
of the pairwise interaction potential~\citep[see, e.g.\@, footnote~{5} in][]{Hamilton2021}.
Nonetheless, in Eq.~\eqref{secondorder_Landau_eq},
we note that the cross term, ${ \bk \!\cdot\! \p F_{3} / \p \bJ }$,
does not explicitly involve the interaction potential
and its precise form is key to ensure all the conservation laws
and $H$ theorem of the kinetic equation~\citep{Fouvry2022}.
As a consequence, we expect that the inhomogeneous ${1/N^{2}}$ \BL\ equation
can be obtained from Eq.~\eqref{secondorder_Landau_eq}
through some intricate substitution
${ |\Lambda_{k_{1}k_{2}}(\bJ)|^2 \!\to\! |\Lambda^{\rd}_{k_{1}k_{2}}[F](\bJ)|^2 }$,
which is still unknown.

In Eq.~\eqref{secondorder_Landau_eq},
the three-body cross term, ${ \bk \!\cdot\! \p F_{3} / \p \bJ }$,
never  vanishes at resonance
except for the thermodynamical equilibrium~\citep[see~\S{IV.C} in][]{Fouvry2022}.
In that sense, three-body collisions always involve
non-trivial resonances and cannot generically vanish
whatever the \DF\@:
this is in sharp contrast with the first-order kinetic blocking 
of Eq.~\eqref{firstorder_Landau_eq}.

The goal of further delaying the relaxation described by Eq.~\eqref{secondorder_Landau_eq},
was investigated in~\S{IV.D} of~\cite{Fouvry2022}.
Therein, they showed that the pairwise potential 
from Eq.~\eqref{def_U_B2}
in conjunction with the profile (3) from Eq.~\eqref{freq_3}
ensures that ${ \Lambda_{k_{1} k_{2}} (\bJ) \!=\! 0 }$ at resonance.
In that case, one gets ${ \p F (J) / \p t \!=\! 0 }$
in Eq.~\eqref{secondorder_Landau_eq},
i.e.\ this kinetic equation predicts no relaxation
whatever the considered (stable) ${ F (J) }$.
We call this a second-order \textit{bare} kinetic blocking.

Let us emphasise that the first-order 
blocking of Eq.~\eqref{firstorder_Landau_eq}
relies on the vanishing of the crossed term ${ \bk \!\cdot\! \p F_{2} / \bJ \!=\! 0 }$,
while the second-order blocking of Eq.~\eqref{secondorder_Landau_eq}
relies on the vanishing of the coupling coefficients 
${ \Lambda_{k_{1} k_{2}} (\bJ) \!=\! 0 }$ at resonance.
This is a fundamental difference.
Indeed, the vanishing of the bare coefficients
${ \Lambda_{k_{1} k_{2}} (\bJ) }$,
does not imply the vanishing of the dressed coefficients, ${ \Lambda^{\rd}_{k_{1} k_{2}} [F] (\bJ) }$,
given that such a dressing depends on the considered \DF\@.
Since ${ \Lambda_{k_{1}k_{2}}^{\rd} [F] (\bJ) }$ will generically
be non-zero at resonance, the inhomogeneous ${1/N^{2}}$ \BL\ equation
is not expected to vanish.
We claim that this will prevent any system from ever undergoing a
second-order \textit{full} kinetic blocking.

\subsection{Scalings of the relaxation}
\label{sec:RelaxationScalings}

Let us now detail the scaling of the relaxation time
w.r.t.\ the total number of particles $N$
and the amplitude of the pairwise interaction, $G$,
in these various regimes.

In Eqs.~\eqref{firstorder_Landau_eq} and~\eqref{secondorder_Landau_eq},
the scaling w.r.t.\ $N$ is straightforwardly read
from the dependence w.r.t.\ the individual mass ${ \mu \!=\! \Mtot / N }$.
One has ${ \p F / \p t \!\propto\! 1/N}$ [resp.\ ${ \!\propto\! 1/N^{2} }$]
in Eq.~\eqref{firstorder_Landau_eq} [resp.\ Eq.~\eqref{secondorder_Landau_eq}].
In the present dynamically hot limit,
the scaling w.r.t.\ $G$ stems from the scaling of the coupling
coefficients.
In Eq.~\eqref{firstorder_Landau_eq},
one has ${ U_{k} \!\propto\! G }$,
so that ${ \p F / \p t \!\propto\! G^{2} }$.
As for Eq.~\eqref{secondorder_Landau_eq},
the bare coupling coefficients, ${ \Lambda (\bJ) }$,
are quadratic in the interaction potential,
i.e.\ ${ \Lambda (\bJ) \!\propto\! G^{2} }$ (see Appendix~\ref{app:1N2Landau}).
As a consequence, Eq.~\eqref{secondorder_Landau_eq},
leads to ${ \p F / \p t \!\propto\! G^{4} }$.
To summarise, in the dynamically hot limit,
the ${1/N}$ Landau Eq.~\eqref{firstorder_Landau_eq}
predicts a relaxation timescale
of order ${ \Trelax \!\propto\! \Tdyn N / G^{2} }$.
And, in the same hot limit,
the ${ 1/N^{2} }$ Landau Eq.~\eqref{secondorder_Landau_eq}
predicts a relaxation on the (slower) timescale
${ \Trelax \!\propto\! \Tdyn N^{2} / G^{4} }$.
In both cases, the larger the number of particles, the slower the evolution;
the stronger the interaction, the faster the evolution.

Now, we need to consider the case of systems
subject to a second-order bare kinetic blocking.
In the dynamically hot limit,
i.e.\ for ${ G \!\to\! 0 }$,
one expects for the ${1/N^{2}}$ dressed coefficients,
${ \Lambda^{\rd} [F] (\bJ) }$,
to converge to the bare ones, ${ \Lambda (\bJ) }$.
Since ${ \Lambda (\bJ) \!\propto\! G^{2} }$,
this leads to an expansion of the form
\begin{equation}
    \label{Lambda_BL_G0}
    \Lambda^{\rm d}[F](\bJ) \underset{G\to 0}{=}
    \Lambda(\bJ) + G^3 \Lambda^{\rm d}_{(3)}[F](\bJ) + {\cal O}(G^4).
\end{equation}
For systems undergoing a second-order bare kinetic blocking,
one has ${ \Lambda (\bJ) \!=\! 0 }$ at resonance.
As a consequence,
in the hot limit,
one finds the asymptotic scaling
${ \Lambda^{\rd} [F] (\bJ) \!\propto\! G^{3} }$.
Given that ${ \p F / \p t \!\propto\! |\Lambda^{\rd} [F] (\bJ)|^{2} }$,
systems subject to a second-order bare kinetic blocking
are therefore expected to relax on a timescale of order
${ \Trelax \!\propto\! \Tdyn N^{2} / G^{6} }$.
In that limit, relaxation is driven by ``leaks''
from dressed three-body interactions.
Phrased differently, ${1/N^{2}}$ effects,
albeit made less efficient by a second-order bare kinetic blocking,
are always driving some non-zero relaxation in
the present long-range interacting inhomogeneous ${1D}$ systems.
We claim that one cannot design
a system in which three-body correlations
would systematically drive no dynamics, whatever the 
considered \DF\@.

One could be worried that four-body correlations,
i.e.\ ${ 1/N^{3} }$ effects, could drive relaxation
more efficiently than the previous leaks
from three-body collective effects.
In Appendix~\ref{app:LandauEquationNNE},
placing ourselves in the hot limit,
we justify that ${ 1/N^{3} }$ effects
drive relaxation on a timescale
of order ${ \Trelax \!\propto\! \Tdyn \, N^{3}/G^{6} }$,
i.e.\ a subdominant process.
As a conclusion, even if it was derived,
an inhomogeneous ${1/N^{3}}$ Landau equation
can never be the main driver of relaxation
in the asymptotic limit ${ N \!\gg\! 1 }$.
This is one of the main results of the present investigation.

\section{Numerical measurements}
\label{sec:NumericalMeasurements}

Let us now recap for each of the frequency profiles
considered in Eq.~\eqref{freq_profile},
the scaling of the relaxation time
expected as one varies
the total number of particles, $N$,
and the amplitude of the pairwise coupling, $G$.
We recall that we place ourselves within the limit of a dynamically hot system,
i.e.\ ${ G \!\to\! 0 }$.
\begin{itemize}
\item \textbf{Profile (1)}. This profile is non-monotonic.
This allows for non-local resonances, ${ J_{1} \!\neq\! J }$,
in the ${1/N}$ Landau Eq.~\eqref{firstorder_Landau_eq}.
The system is not subject to any kinetic blocking.
We expect then ${ \Trelax \!\propto\! \Tdyn N/G^{2} }$.
\item \textbf{Profile (2)}. This profile is monotonic,
so that the ${1/N}$ Landau and \BL\ operators both vanish.
The ${1/N}$ dynamics is \textit{fully} blocked.
The system can only relax through ${1/N^{2}}$
effects, as governed by Eq.~\eqref{secondorder_Landau_eq}
in the hot regime.
The profile (2) is not submitted to any second-order bare kinetic blocking,
i.e.\ Eq.~\eqref{secondorder_Landau_eq} gives a non-vanishing contribution.
We expect then the scaling ${ \Trelax \!\propto\! \Tdyn N^{2}/G^{4} }$.
\item \textbf{Profile (3)}. This profile is monotonic,
hence the ${1/N}$ dynamics is fully blocked. 
In addition, following~\cite{Fouvry2022},
this profile is also submitted to a second-order bare kinetic blocking,
i.e.\ the ${1/N^{2}}$ Eq.~\eqref{secondorder_Landau_eq} vanishes.
Yet, even though the ${1/N^{2}}$ Landau equation is zero
whatever the considered \DF\@,
we argued in Sec.~\ref{sec:RelaxationScalings} that leaks from the
(yet unknown) ${1/N^{2}}$ \BL\ equation
will lead to a relaxation time scaling like ${ \Trelax \!\propto\! \Tdyn N^{2}/G^{6} }$
and not like ${ \Trelax \!\propto\! \Tdyn N^{3}/G^{6} }$ as 
one could have (wrongly) guessed from four-body correlations 
contribution (see Appendix~\ref{app:LandauEquationNNE}).
\end{itemize}
To summarise, the profiles from Eq.~\eqref{freq_profile}
are predicted to be associated with relaxation times
scaling like
\begin{subequations}
\begin{align}
{} & \textbf{Relaxation times}
\nonumber
\\
{} & \quad\quad\quad \text{(1):} \;\;\; \Trelax \propto \Tdyn N/G^{2} ;
\label{Trelax_1}
\\
{} & \quad\quad\quad \text{(2):} \;\;\; \Trelax \propto \Tdyn N^{2}/G^{4} ;
\label{Trelax_2}
\\
{} & \quad\quad\quad \text{(3):} \;\;\; \Trelax \propto \Tdyn N^{2}/G^{6} .
\label{Trelax_3}
\end{align}
\label{Trelax_123}\end{subequations}

We set out to recover numerically the scalings
predicted in Eq.~\eqref{Trelax_123}.
To do so, for a given frequency profile,
we explore a ${ 5 \!\times\! 5 }$ grid of values of ${ (N,G) }$
running for each a large batch of $10^{3}$ simulations.
We then estimate from these the dependence
of the relaxation time w.r.t.\ ${ (N,G) }$.
We refer to Appendix~\ref{app:NBody} for details
on how dynamics driven by Eq.~\eqref{def_U_B2} may be efficiently
integrated.

In practice, we search for a power-law
dependence of the form
\begin{equation}
\Trelax \propto \Tdyn \, N^{\gamma_{N}} / G^{\gamma_{G}} ,
\label{ansatz_Trelax}
\end{equation}
and constraints on the power indices ${ (\gamma_{N},\gamma_{G}) }$.
A handful of reasons make these measurements challenging.

First, one needs to ensure that $G$ is small enough
for collective effects to be negligible,
though large enough so that relaxation
can still be observed numerically.
We rely on linear response theory (Appendix~\ref{app:LinearResponse})
to determine appropriate ranges in $G$.

Second, long-time integrations of the model
from Eq.~\eqref{def_U_B2} are challenging,
given the limited smoothness of this interaction potential.
To ease this exploration, we accelerate the evaluation of the forces using an (exact)
multipole method, as detailed in Appendix~\ref{app:Multipole}.
We also carefully pick our integration parameters
to keep the global integration errors under control (Appendix~\ref{app:ErrorsIntegration}).

Finally, since we consider simulations
with small number of particles, the estimation
of the relaxation time requires some care
to prevent possible biases (Appendix~\ref{app:Measurements}).

\begin{figure}[htbp!]
    \begin{center}
    \includegraphics[width=0.45\textwidth]{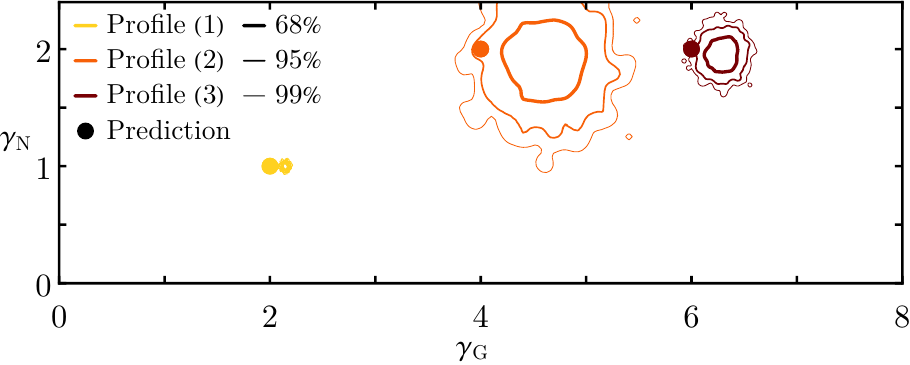}
    \caption{Dependence of the relaxation time,
via the power-law indices ${(\gamma_{N}, \gamma_{G})}$ from Eq.~\eqref{ansatz_Trelax},
as a function of the total number of particles, $N$,
and the strength of the pairwise interaction, $G$,
for the various frequency profiles from Eq.~\eqref{freq_profile}.
As explained in Appendix~\ref{app:Bias}, a systematic bias 
towards higher ${\gamma_{G}}$ is to be expected.
    \label{fig:Scaling}}
    \end{center}
\end{figure}

In Fig.~\ref{fig:Scaling}, we report our main result
namely the measurement of the power indices ${ (\gamma_{N},\gamma_{G}) }$
from Eq.~\eqref{ansatz_Trelax}
as one varies the considered frequency profiles.

In that figure, we first recover that all profiles
exhibit their expected scaling w.r.t.\ $N$,
i.e.\ the value of $\gamma_{N}$.
In particular, even though the profile (3)
is submitted to a second-order bare kinetic blocking,
i.e.\ Eq.~\eqref{secondorder_Landau_eq} exactly vanishes,
its relaxation is still driven by ${ 1/N^{2} }$ effects,
i.e.\ three-body correlations.
Once again, we emphasise that this particularly slow relaxation
is sourced by leaks from dressed three-body correlations
and not by four-body correlations.

In Fig.~\ref{fig:Scaling}, we also find that all profiles
show scalings w.r.t.\ $G$, i.e.\ the value of $\gamma_{G}$,
in agreement with the predictions.
Though, one could be suspicious about the systematic bias
in the value of $\gamma_{G}$, which is always measured
to be larger than the predicted one.
We argue that this was to be expected since
the measurements were made for a finite value of $G$,
while the predictions correspond to the limit ${ G \!\to\! 0 }$.
In particular, given the difficulty of integrating the motion
driven by the potential from Eq.~\eqref{def_U_B2}
(see Appendix~\ref{app:TimeIntegration}), we had to limit ourselves to considering
not so dynamically hot systems (see Table~\ref{table:B2}).
In Appendix~\ref{app:Bias}, we show that the bias observed
in Fig.~\ref{fig:Scaling} is well within the limits
that could be expected
from the effective use of finite values of $G$.

Ultimately, Fig.~\ref{fig:Scaling} allows us
to clearly confirm numerically the scaling of the relaxation
times predicted in Eq.~\eqref{Trelax_123},
along with all the signatures associated with these various
kinetic blockings.

\section{Conclusions}
\label{sec:Conclusion}

In this paper,
we investigated the long-term evolution
of long-range interacting
inhomogeneous ${1D}$ systems.
In particular, we highlighted the existence of two types
of kinetic blockings:
(i) a first-order \textit{full} kinetic blocking
in systems with
a monotonic frequency profile and subject only ${1\!:\!1}$ resonances,
associated with the vanishing of the
inhomogeneous ${1/N}$ \BL\ equation;
(ii) a second-order \textit{bare} kinetic blocking
associated with the vanishing
of the inhomogeneous ${1/N^{2}}$ Landau equation.

Considering a fixed interaction potential (Eq.~\ref{def_U_B2}),
we presented a large numerical exploration
to confirm the existence of these various blockings.
In particular, we showed that dynamically hot systems submitted to a second-order bare
kinetic blocking still relax via ${1/N^2}$ effects,
as a result of ``leaks'' from collective effects.
We argued that the (still unknown)
${1/N^2}$ \BL\@ collision operator would never vanish
as the cancellation of the ${ 1/N^{2} }$ Landau operator 
arises from very specific conditions
on the coupling terms,
which change when taking collective effects into account.
Hence, four-body correlations,
i.e.\ ${1/N^{3}}$ effects, can never be the main driver
of the relaxation of long-range interacting inhomogeneous systems,
even in the present contrived ${1D}$ geometry.
As such, no system will ever suffer from a 
second-order \textit{full} kinetic blocking, and 
we claim that the ${1/N^3}$ Landau equation does not 
need further investigation.

The present numerical work is only one more step towards
a finer understanding of (very) long-term dynamics
and high-order correlations.
Naturally, it would be worthwhile and rewarding
to derive the ${1/N^2}$ \BL\ equation,
hence generalising Eq.~\eqref{secondorder_Landau_eq}. This is no easy task,
and we feel that a realistic roadmap
would be to perform such a delicate calculation
first for the (single-harmonic) homogeneous Hamiltonian Mean Field model~\cite{Rocha+2014,Fouvry+2019},
then for homogeneous systems
with an arbitrary potential of interaction~\citep{Fouvry+2020},
and ultimately for the present inhomogeneous regime.
Ultimately, this line of work would convincingly show
that second-order kinetic blockings
can never lead to the full vanishing of ${1/N^{2}}$ effects.

\begin{acknowledgments}
This work is partially supported by the grant Segal ANR-19-CE31-0017 
of the French Agence Nationale de la Recherche
and by the Idex Sorbonne Universit\'e.
We warmly thank S.\ Rouberol for the smooth running
of the Infinity cluster
where the simulations were performed.
We warmly thank C.\ Pichon for stimulating discussions.
\end{acknowledgments}

\vskip 1cm

\appendix

\section{Linear response theory}
\label{app:LinearResponse}

In this section, we rely on linear response theory~\citep[see, e.g.\@,~\S{5.3} in][]{Binney+2008}
to estimate a system's dynamical temperature.
This allows us then to carefully pick the value of $G$
to place ourselves within the dynamically hot limit.
Because the interaction potential from Eq.~\eqref{def_U_B2}
is neither fully attractive nor repulsive,
we first need to (slightly) generalise the computation of the response matrix,
compared to the typical gravitational case.

\subsection{Basis elements}
\label{app:BasisElements}

Let us consider a unidimensional Hamiltonian system,
with the angle-action coordinates ${ \bw \!=\! (\theta , J) }$.
Taking inspiration from~\cite{Kalnajs1976},
we define a set of potential/density basis elements
${(\psi^{(p)}, \rho^{(p)})}$ via
\begin{subequations}
    \begin{align}
        {} & \psi^{(p)} (\bw) = \!\! \int \!\! \rd \bwp \, \rho^{(p)} (\bwp) \, U (\bw , \bwp) ,
        \label{Poisson_basis}
        \\
        {} & \!\! \int \!\! \rd \bw \, \rho^{(p)} (\bw) \, \psi^{(q) *} (\bwp) = - \gamma_{p} \, \delta_{p}^{q} ,
        \label{norm_basis}
    \end{align}
    \label{def_basis}\end{subequations}
where ${ \gamma_{p} \!=\! \pm 1 }$.
With such a convention, the case ${ \gamma_{p} \!=\! + 1}$
[resp.\ ${- 1}$] corresponds, typically,
to the attractive case [resp.\ repulsive].
Using Eq.~\eqref{def_basis},
one can decompose the interaction potential as
\begin{equation}
    U (\bw , \bwp) = - \sum_{p} \psi^{(p)} (\bw) \, \Gamma_{pq} \, \psi^{(q) *} (\bwp) ,
    \label{decomp_U}
\end{equation}
with the diagonal matrix ${\Gamma_{pq} \!=\! \gamma_{p} \delta_{p}^{q} }$.
The usual self-gravitating case corresponds to ${ \bGamma \!=\! \bI }$,
the identity matrix.

\subsection{Response matrix}
\label{app:ResponseMatrix}

Following Kalnajs' matrix method \citep{Kalnajs1976} and
paying particular attention to the 
basis normalisation \citep[see, e.g.\@,][]{Dootson+2022},
the susceptibility of the self-gravitating system
is captured by the dielectric matrix
\begin{equation}
    \bE (\omega) = \bGamma - \bM (\omega) ,
    \label{def_bE}
\end{equation}
with the response matrix
\begin{equation}
    M_{pq} (\omega) = 
    - 2 \pi \sum_{k} \!\! \intL \!\! \rd J \, 
    \frac{k \, \p F / \p J}{k\,\Omega(J) \!-\! \omega} \, 
    \psi^{(p) *}_{k} (J) \, 
    \psi^{(q)}_{k} (J) .
    \label{def_bM}
\end{equation}
In this expression, we used 
the Fourier transform of the basis elements
given by 
\begin{equation}
    \psi^{(p)}_{k} (J) = \!\!\int\! \frac{\rd\theta}{2\pi}\, \psi^{(p)} (\bw) \, \re^{-\ri k \theta}.
    \label{exp_basis_FT}
\end{equation}
We refer to~\cite{FouvryPrunet2022} and references therein
for a discussion
on the resonant denominator, ${ 1/(\bk \!\cdot\! \bO (\bJ) \!-\! \omega) }$,
and the associated Landau integral, ${ \int_{\mL} \rd \bJ }$.

We stress that the response matrix scales as ${\bM(\omega)\!\propto\!G}$
since the potential basis elements
(and therefore their Fourier transform) scale like ${\psi^{(p)}\!\propto\!\sqrt{G}}$.
Consequently, in the dynamically hot limit,
i.e.\ ${ G \!\to\! 0 }$,
one has ${\bE(\omega)\!\to\!\bGamma}$.
The smaller $G$, the smaller the collective effects.

\subsection{Natural basis elements}
\label{app:B2BasisElements}

We now focus on the interaction potential
from Eq.~\eqref{def_U_B2}. We aim at constructing
natural basis elements for it, following Eq.~\eqref{def_basis}.
Equation~\eqref{def_U_B2} can be rewritten as
\begin{align}
U(\bw,\bwp) {} & = G \, (J \!-\! \Jp)^{2} \, \mB_{2} [\theta \!-\! \thetap]
\nonumber
\\
{} & =  G \, U_{J} (J,\Jp) \, U_{\theta} (\theta , \thetap) .
\label{rewrite_B2}
\end{align}

We decompose the angular part of the potential with
\begin{align}
U_{\theta} (\theta , \thetap) {} & = \mB_{2} [\theta \!-\! \theta_{2}]
\nonumber
\\
{} & = \frac{1}{\pi^{2}} \sum_{k = 1}^{+ \infty} \frac{1}{k^{2}} \cos [k (\theta \!-\! \thetap)]
\nonumber
\\
{} & = \sum_{\mathclap{p_{\theta} \neq 0}} \frac{1}{2 \pi^{2}} \frac{1}{p_{\theta}^{2}} \, \re^{\ri p_{\theta} \theta} \, \re^{- \ri p_{\theta} \thetap} .
\label{decomp_Utheta}
\end{align}

Following an approach similar to~\S{A2} in~\cite{Roule+2022},
we periodise the function ${ U(J,\Jp) }$
on a period ${2 \JL}$.
More precisely, we define ${ \Uper (J , \Jp) = U (J,\Jp) }$
for ${ |J \!-\! \Jp| \!\leq\! \JL }$
and ${ \Uper (J \!+\! 2 k \JL,\Jp) \!=\! \Uper (J,\Jp) }$
for ${ k \!\in\! \mathbb{Z} }$.
Then, we have the Fourier decomposition
\begin{align}
\Uper (J , \Jp) {} & = \tfrac{1}{3} \JL^{2} + \sum_{\mathclap{p_{J} = 1}}^{+ \infty} \frac{4 (-1)^{p_{J}}}{\pi^{2} p_{J}^{2}} \JL^{2} \cos [p_{J} \tfrac{\pi}{\JL} (J \!-\! \Jp)]
\nonumber
\\
{} & = \tfrac{1}{3} \JL^{2} + \sum_{\mathclap{p_{J} \neq 0}} \frac{2 (-1)^{p_{J}}}{\pi^{2} p_{J}^{2}} \JL^{2} \re^{\ri p_{J} \frac{\pi}{\JL} J} \, \re^{- \ri p_{J} \frac{\pi}{\JL} \Jp} .
\label{decomp_UJ}
\end{align}

We are now set to define our basis elements.
We index them with ${ p \!=\! (p_{\theta} , p_{J}) }$
(with ${ p_{\theta} \!\neq\! 0 }$), and introduce
\begin{equation}
\psi^{(p)} (\bw) = \sqrt{|G|} \, a_{\theta} (p_{\theta}) \, a_{J} (p_{J}) \, \re^{\ri p_{\theta} \theta} \, \re^{\ri p_{J} \frac{\pi}{\JL} J} .
\label{def_psi_B2}
\end{equation}
In that expression, the (positive) coefficients ${ a_{\theta} (p_{\theta}) }$
and ${ a_{J} (p_{J}) }$ follow from Eqs.~\eqref{decomp_Utheta} and~\eqref{decomp_UJ} and read
\begin{equation}
a_{\theta} (p_{\theta}) = 1/ (\sqrt{2} \, \pi |p_{\theta}|) ,
\label{def_atheta}
\end{equation}
as well as
\begin{equation}
a_{J} (p_{J}) =
\begin{cases}
\displaystyle \JL / \sqrt{3} {} & \text{if} \;\; p_{J} = 0 ,
\\[1.0ex]
\displaystyle \sqrt{2} \JL / (\pi |p_{J}|) {} & \text{otherwise} .
\end{cases}
\label{def_aJ}
\end{equation}

Following the convention from Eq.~\eqref{def_basis}
(and reducing the action integration range to ${ -\JL \!\leq\! J \!\leq\! \JL }$),
one finds that the associated density elements read
\begin{equation}
\rho^{(p)} (\bw) = \frac{(-1)^{p_{J}}}{4 \pi \JL} \, \frac{\Sign[G]}{\sqrt{|G|}} \, \frac{\re^{\ri p_{\theta} \theta} \re^{\ri p_{J} \frac{\pi}{\JL}J}}{a_{\theta}(p_{\theta}) \, a_{J} (p_{J})} .
\label{def_rho_B2}
\end{equation}
Finally, one finds the normalisation constant
\begin{equation}
\gamma_{p} = - (-1)^{p_{J}} \Sign[G] .
\label{val_eps_B2}
\end{equation}

\subsection{Computing the response matrix}
\label{app:CompbM}

Using the basis elements from Eq.~\eqref{def_psi_B2},
whose Fourier transform satisfies
${\psi^{(p)}_{k}\!\propto\!\delta^{p_{\theta}}_{k}}$,
we rewrite the response matrix from Eq.~\eqref{def_bM} as
\begin{equation}
M_{pq} (\omega) = \delta_{p_{\theta}}^{q_{\theta}} \, A \, \!\! \intL \!\! \rd J \, \frac{g (J)}{h (J) \!-\! \omega} , 
\label{calc_M}
\end{equation}
with
\begin{subequations}
\begin{align}
A {} & = - 2 \pi |G| \, a_{\theta}^{2} [p_{\theta}] \, a_{J} [p_{J}] \, a_{J} [q_{J}] p_{\theta} ,
\label{def_A_M}
\\
g (J) {} & =  \frac{\p F}{\p J} \, \re^{- \ri (p_{J} - q_{J}) \tfrac{\pi}{\JL} J} ,
\label{def_g_M}
\\
h (J) {} & = p_{\theta} \, \Omega (J) .
\label{def_h_M}
\end{align}
\label{def_Agh_M}\end{subequations}

To carry out the integral in Eq.~\eqref{calc_M},
we truncate the domain ${ - \JL \!\leq\! J \!\leq\! \JL }$
into $\KJ$ uniform intervals of length ${ \Delta J \!=\! 2 \JL / \KJ }$
centred around the locations
${ J_{k} \!=\! - \JL \!+\! (k \!+\! \tfrac{1}{2}) \Delta J }$.
Equation~\eqref{calc_M} becomes
\begin{equation}
M_{pq} (\omega) = \delta_{p_{\theta}}^{q_{\theta}} \, A \, \sum_{k = 1}^{\KJ} I_{k} (\omega) ,
\label{calc_M_disc}
\end{equation}
with the integrals
\begin{align}
I_{k} (\omega) {} & = \!\! \intLA \!\!\!\!\!\! \rd J \, \frac{g (J)}{h(J) \!-\! \omega}
\nonumber
\\
{} & \simeq \!\! \intLB \!\!\!\!\!\! \rd \delta J \, \frac{g_{0} + g_{1} \delta J}{h_{0} \!+\! h_{1} \delta J \!-\! \omega} ,
\label{calc_I_M}
\end{align}
using a first-order expansion
with ${ g_{0} \!=\! g (J_{k}) }$, ${ g_{1} \!=\!g^{\prime} (J_{k}) }$,
and a similar notation for ${ (h_{0},h_{1}) }$.

The final step of the calculation
is to compute
\begin{equation}
I_{k} (\omega) \simeq  \!\! \intLb \!\! \rd u \, \frac{\og_{0} \!+\! \og_{1} u}{u \!-\! \varpi} ,
\label{shape_int_I}
\end{equation}
with ${ \og_{0} \!=\! g_{0}/|h_{1}| }$,
${ \og_{1} \!=\! g_{1} \Delta J/(2 h_{1}) }$
and the rescaled frequency
${ \varpi \!=\! 2(\omega \!-\! h_{0})/(|h_{1}|\Delta J) }$.
We recall that the resonant denominator in that expression
has to be interpreted using Landau's prescription~\citep[see, e.g.,][]{Binney+2008}.
In practice, we use the expression presented in~\S{D} of~\cite{FouvryPrunet2022}, evaluated for a purely real frequencies.
The computation presented in Appendix~\ref{app:Nyquist}
used ${ \JL \!=\! 5 }$,
${ \KJ \!=\! 10^{4} }$,
${ p_{\theta} \!=\! 1 }$,
and ${ |p_{J}| \!\leq\! 100 }$.

\subsection{Nyquist contours}
\label{app:Nyquist}

Owing to the Kronecker symbol ${ \delta_{p_{\theta}}^{q_{\theta}} }$
in Eq.~\eqref{calc_M},
${ \bE (\omega) }$, as defined in Eq.~\eqref{def_bE},
is diagonal w.r.t.\
the angular index ${ p_{\theta} }$,
so that the $p_{\theta}$ can be considered independently of one another.
Moreover, since
${ a_{\theta} [p_{\theta}] \!\propto\! 1/|p_{\theta}| }$
(Eq.~\ref{def_atheta}),
the largest values of the response matrix
are obtained for ${ p_{\theta} \!=\! \pm 1 }$.
Finally, by symmetry, we may limit ourselves to ${ p_{\theta} \!=\! 1 }$
when assessing the system's linear stability.

\begin{figure}[htbp!]
    \begin{center}
\includegraphics[width=0.45\textwidth]{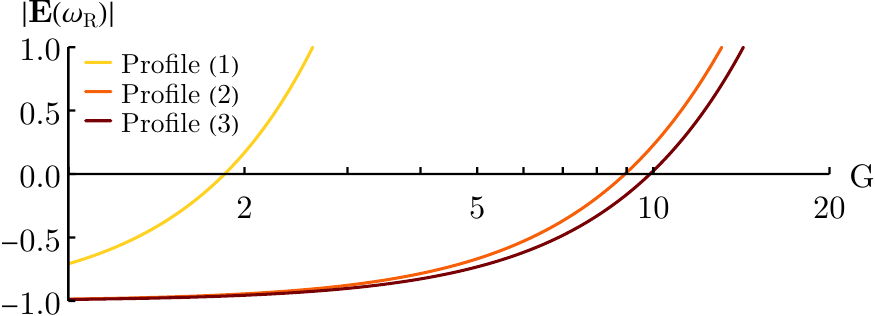}
    \caption{Determinant of the dielectric matrix, ${ |\bE (\omegaRcrit)| }$,
    as a function of $G$, for the profiles from Eq.~\eqref{freq_profile}.
    For the profile (1) [resp.\ (2); (3)],
    the critical frequency is ${ \omegaRcrit \!\simeq\! 1.047 }$
    [resp.\ 0; 0]. The value of $\Gcrit$ is reached when
    ${ |\bE (\omegaRcrit)| }$ crosses zero.
    \label{fig:Gcrit}}
    \end{center}
\end{figure}

For a given profile, we define the critical coupling amplitude,
${ \Gcrit }$, as
\begin{equation}
\Gcrit = \Min \big\{ \, G \!>\! 0 \, \big| \, \exists \, \omegaR \!\in\! \mathbb{R} \;\; \mathrm{s.t.} \;\;  |\bE (\omegaR)| \!=\! 0 \big\} ,
\label{def_Gcrit}
\end{equation}
with ${ |\bE (\omega)| }$ the complex determinant. Systems with ${ 0 \!\leq\! G \!<\! \Gcrit }$ are linearly stable.
And, the smaller ${ G / \Gcrit }$, the hotter the system
and the more negligible collective effects are.
In Fig.~\ref{fig:Gcrit}, we represent the dependence
of ${ G \!\mapsto\! |\bE (\omegaRcrit)| }$,
where the (real) frequency, $\omegaRcrit$,
is the frequency of the system's first oscillating mode.
It satisfies ${ \ImPart [|\bE (\omegaRcrit)|] \!=\! 0 }$
for all $G$ as well as
${ |\bE (\omegaRcrit)| \!=\! 0 }$ for ${ G \!=\! \Gcrit }$.
From that figure, we readily determine the values
of $\Gcrit$ for each of profiles,
as reported in Table~\ref{table:B2}.

\begin{figure}[htbp!]
    \begin{center}
\includegraphics[width=0.45\textwidth]{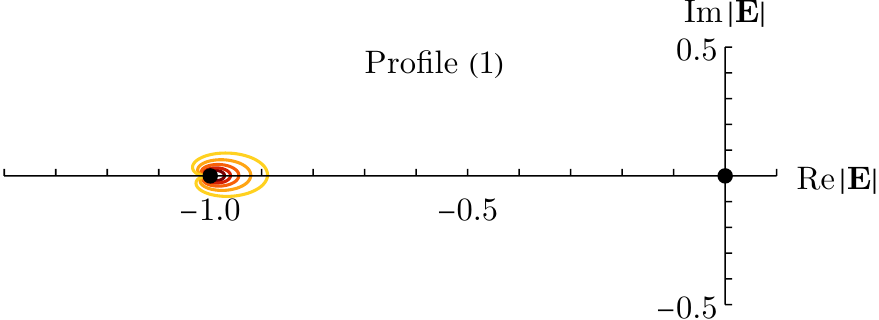}
\includegraphics[width=0.45\textwidth]{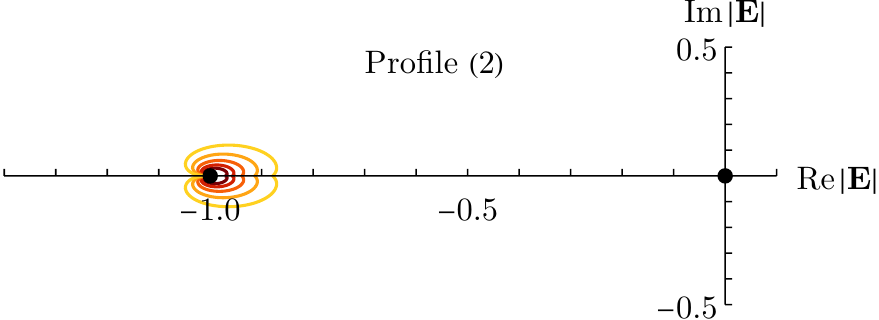}
\includegraphics[width=0.45\textwidth]{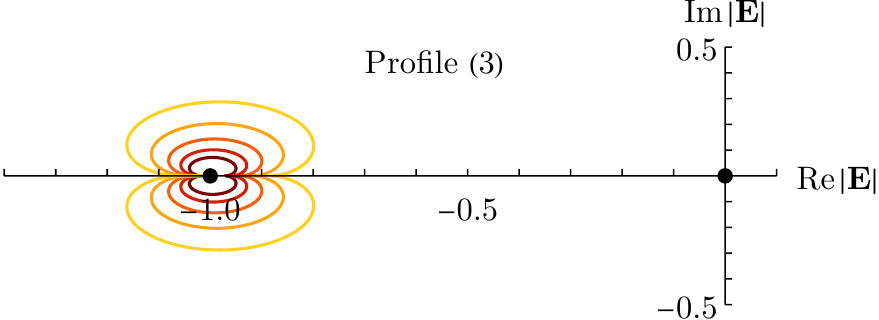}
    \caption{Nyquist contours, ${ \omegaR \!\mapsto\! |\bE (\omegaR)| }$,
for ${ p_{\theta} \!=\! 1 }$
and the various frequency profiles from Eq.~\eqref{freq_profile},
as one increases $G$ (red to yellow)
for the values presented in Table~\ref{table:B2}.
    \label{fig:Nyquist}}
    \end{center}
\end{figure}

In Fig.~\ref{fig:Nyquist}, we illustrate the Nyquist contours~\citep[see, e.g.\@,][]{Chavanis+2009},
${ \omegaR \!\mapsto\! |\bE (\omegaR)| }$,
for the three frequency profiles,
as one varies the coupling amplitude, $G$.
Following Eq.~\eqref{exp_Ud},
the closer ${ \bE (\omega) }$ is from
the identity-like matrix $\bGamma$,
the closer the dressed coupling coefficients
${ \Ud_{k} }$ are from the bare ones
${ U_{k} }$,
i.e.\ the more negligible are collective effects.
Phrased differently, the smaller $G$, the closer is the contour
from the point ${ \det \bGamma \!=\! (-1,0)}$,
i.e.\ the hotter are the systems.
In practice, the ranges of $G$ effectively simulated
are detailed in Table~\ref{table:B2}.

\section{${1/N}$ kinetic equations}
\label{app:1NKE}

Following the notations from Sec.~\ref{sec:1stKinetic},
the inhomogeneous ${1/N}$ Landau equation
generically reads~\cite{Chavanis2013}
\begin{align}
    \frac{\p F(J)}{\p t} = 2\pi \mu \frac{\p }{\p J} \bigg[ \!\sum_{k_1} k_1 {} &
    \!\! \int \!\! \rd J_{1} \, \big| U_{k_1} (\bJ ) \big|^{2} \,  
    \label{exp_1N_Landau_eq}
    \\
    \times {} & \deltaD\big( \bk \!\cdot\! \bO\big) \, 
    \bk\!\cdot\!\frac{\p }{\p \bJ}F_2(\bJ) \bigg].
    \nonumber
\end{align}
In this equation, the bare coupling coefficients,
${ U_{k} (J,\Jp) }$,
are the Fourier transform w.r.t.\ the angles ${ (\theta,\thetap) }$
of the interaction potential, ${ U(|\theta-\thetap|,\{J,\Jp\}) }$.
They read
\begin{equation}
    U (\bw , \bwp) = \sum_{\mathclap{k = - \infty}}^{\mathclap{+ \infty}} 
    U_{k} (J,\Jp) \, \re^{\ri k (\theta - \thetap)} ,
    \label{def_Uk}
\end{equation}
with
\begin{equation}
    U_{k} (J , \Jp) = \!\! \int \!\! \frac{\rd \theta}{2 \pi} 
    \frac{\rd \thetap}{2 \pi} \, 
    U \big( \bw , \bwp \big) \, 
    \re^{- \ri k( \theta - \thetap)} .
    \label{exp_Uk}
\end{equation}
Following Eqs.~\eqref{decomp_U} and~\eqref{exp_basis_FT},
the bare coupling coefficients can equivalently
be rewritten using the basis elements,
to become
\begin{equation}
    U_{k} (J,\Jp) = - \sum_{p,q} \psi^{(p)}_{k} (J) 
    \, \Gamma_{pq} \, \psi^{(q) *}_{k} (\Jp).
    \label{exp_Uk_basis}
\end{equation}
For an interaction potential of typical amplitude $G$,
one has ${ U_{k} \!\propto\! G }$,
so that the relaxation sourced by Eq.~\eqref{exp_1N_Landau_eq}
occurs on a timescale of order ${ \Trelax \!\propto\! \Tdyn N / G^{2} }$.

When accounting for collective effects,
the ${1/N}$ Landau equation becomes the ${1/N}$ \BL\ equation~\citep{Heyvaerts2010,Chavanis2012}.
It is straightforwardly obtained from Eq.~\eqref{exp_1N_Landau_eq}
by replacing the bare coupling coefficients
with their (frequency-dependent) dressed analogues 
${ \Ud_{k} (J,\Jp,\omega) }$.
Using the self-consistency relation from Eq.~{(F5)} in~\cite{Fouvry+2018},
they read
\begin{equation}
    \Ud_{k} (J , \Jp , \omega) = 
    - \sum_{p,q} \psi^{(p)}_{k} (J) \, 
    E_{pq}^{-1} (\omega) \, \psi^{(q) *}_{k} (\Jp),
    \label{exp_Ud}
\end{equation}
with $\bE$ the dielectric matrix from Eq.~\eqref{def_bE}.
In practice, in the ${1/N}$ \BL\ equation,
the dressed coupling coefficients are evaluated at the 
resonant frequency ${ \omega \!=\! k \,\Omega(J) }$.
Accordingly, in Sec.~\ref{sec:1stKinetic},
we shorten the notations and write
${\Ud_{k} (J , \Jp) \!=\! \Ud_{k} (J , \Jp, k\,\Omega[J])} $.

\section{${1/N^2}$ Landau equation}
\label{app:1N2Landau}

We reproduce here the inhomogeneous ${1/N^{2}}$ Landau
equation presented in Eq.~{(4)} of~\cite{Fouvry2022}.
Using the notations from Sec.~\ref{sec:2ndKinetic}, it reads
\begin{align}
\frac{\p F (J)}{\p t} = 2 \pi^{3} \mu^{2} \frac{\p }{\p J} 
\bigg[ {} & \sum_{k_{1} , k_{2}} \!(k_{1} \!+\! k_{2}) 
\mP \!\!\! \int \!\! \rd J_{1} \rd J_{2}\, |\Lambda_{k_{1}k_{2}} (\bJ)|^2 
\nonumber
\\
\times {} &
\deltaD \big[ \bk \!\cdot\! \bO \big] \, \bigg( \bk \!\cdot\! \frac{\p }{\p \bJ} \bigg) \, F_{3} (\bJ) \bigg].
\label{exp_Landau_1N2}
\end{align}
In that expression, the sum over the resonance vectors ${ (k_{1}, k_{2}) }$
is restricted to those such that
$k_{1}$, $k_{2}$, and ${ k_{1} \!+\! k_{2} }$
are all non-zeros.
Equation~\eqref{exp_Landau_1N2}
also involves $\mP$, Cauchy principal value,
acting on the integral 
over ${ \rd J_{1} }$,
and we refer to~\S{C2} in~\cite{Fouvry2022}
to prove the well-posedness of the associated high-order
resonant denominator.
Finally, the coupling coefficients, ${ \Lambda_{k_{1}k_{2}} (\bJ) }$,
read~\citep[see~\S{B} in][]{Fouvry2022}
\begin{equation}
    \Lambda_{k_{1}k_{2}} (\bJ) \!=\! \frac{
    \big[\Omega(J) \!-\! \Omega(J_{1})\big] \, 
    \mU_{k_{1}k_{2}}^{(1)} (\bJ) + 
    k_{2} \, \mU^{(2)}_{k_{1}k_{2}} (\bJ)}{
    k_{1} (k_{1} \!+\! k_{2}) \big[\Omega(J) \!-\! \Omega(J_{1})\big]^2 } ,
    \label{def_Lambda}
\end{equation}
with the coupling functions $\mU^{(1,2)}$
defined in the same section.
Equation~\eqref{exp_Landau_1N2} conserves 
the total mass, energy, and momentum,
and satisfies an $H$-theorem \cite{Fouvry2022}.
Importantly, for an interaction potential of amplitude $G$,
one has ${ \mU^{(1,2)}_{k_{1}k_{2}} (\bJ) \!\propto\! G^{2} }$.
The relaxation sourced by Eq.~\eqref{exp_Landau_1N2}
occurs therefore on a timescale of order ${ \Trelax\!\propto\!\Tdyn N^2/G^4 }$.

\section{${1/N^3}$ Landau equation}
\label{app:LandauEquationNNE}

Our goal in this Appendix is not to derive a detailed kinetic equation
but rather to investigate the scaling w.r.t.\ $N$ and $G$
of the resulting evolution,
when driven by ${1/N^{3}}$ effects in the hot limit.
We follow~\S{A1} of~\cite{Fouvry2022}
for the setup of the notations.
The dynamics of a $N$-body system is exactly described
by the BBGKY equations for the $n$-body \DFs\@,
${ F_{n} \!=\! F_{n} (\bw_{1}, ... , \bw_{n} , t) }$.
They read
\begin{equation}
\frac{\p F_{n}}{\p t} \!+\! \big[ F_{n} , H_{n} \big]_{n} \!+\! \!\! \int \!\! \rd \bw_{n+1} \big[ F_{n+1} , \delta H_{n+1} \big]_{n+1} \!= 0 ,
\label{BBGKY_Fn}
\end{equation}
with the Poisson bracket
\begin{equation}
\big[ f, h \big]_{n} = \sum_{i = 1}^{n} \bigg( \frac{\p f}{\p \theta_{i}} \frac{\p h}{\p J_{i}} - \frac{\p f}{\p J_{i}} \frac{\p h}{\p \theta_{i}} \bigg) .
\label{def_Poisson}
\end{equation}
We also introduced the specific Hamiltonian
\begin{equation}
H_{n} = \sum_{i = 1}^{n} \Uext (\bw_{i}) + \sum_{i < j}^{N} \mu \, U (\bw_{i} , \bw_{j}) ,
\label{def_Hn}
\end{equation}
as well as the specific interaction energy
\begin{equation}
\delta H_{n+1} (\bw_{1} , ... , \bw_{n+1}) = \sum_{i = 1}^{n} U (\bw_{i} , \bw_{n+1}) .
\label{def_deltaH}
\end{equation}
Equation~\eqref{BBGKY_Fn} provides us
with evolution equations for $F_{1}$ up to $F_{4}$.

To perform perturbative expansions w.r.t.\
the small parameter ${1/N}$,
we rely on the cluster expansion~\citep{Balescu1997}.
For the sake of completeness, we reproduce here
explicitly the associated expressions.
For $F_{2}$, we introduce
\begin{align}
F_{2} (1,2) = {} & F (1) \, F (2)
\nonumber
\\
+ {} &  G_{2} (1,2) ,
\label{cluster_F2}
\end{align}
using the shortened notation ${ 1 \!=\! \bw_{1} }$.
Similarly, the three-body \DF\@, $F_{3}$,
is expanded as
\begin{align}
F_{3} (1,2,3) = {} & F(1) \, F(2) \, F(3)
\nonumber
\\
+ {} & \sum_{a = 1}^{3} F (a) \, G_{2} (\rest)
\nonumber
\\
+ {} & G_{3} (1,2,3) ,
\label{cluster_F3}
\end{align}
with
${ \rest \!=\! (2,3) }$ when ${a \!=\! 1}$
and so forth.
Similarly, we decompose $F_{4}$ as
\begin{align}
F_{4} (1,2,3,4) = {} & F (1) \, F (2) \, F(3) \, F(4)
\nonumber
\\
+ {} & \sum_{\mathclap{a = 1}}^{3} \sum_{b = a + 1}^{4} \!\!\! F (a) \, F(b) \, G_{2} (\rest)
\nonumber
\\
+ {} & \sum_{a = 1}^{4} F (a) \, G_{3} (\rest)
\nonumber
\\
+ {} & \sum_{a = 2}^{4} G_{2} (1,a) \,  G_{2} (\rest)
\nonumber
\\
+ {} & G_{4} (1,2,3,4) .
\label{cluster_F4}
\end{align}
Finally, the five-body \DF\@, $F_{5}$, is decomposed as
\begin{align}
F_{5} (1,2,3,4,5) = {} & F(1) \, F(2) \, F(3) \, F(4) \, F(5)
\nonumber
\\
+ {} & \sum_{a = 1}^{3} \sum_{b = a + 1}^{4} \sum_{c = b + 1}^{5} \!\!\! F(a) \, F(b) \, F(c) \, G_{2} (\rest)
\nonumber
\\
+ {} & \sum_{a = 1}^{4} \sum_{b = a + 1}^{5} \!\!\! F(a) \, F(b) \, G_{3} (\rest) 
\nonumber
\\
+ {} & \sum_{a = 1}^{5} \sum_{b = a + 2}^{a + 4} F(a) \, G_{2} (a + 1 , b) \, G_{2} (\rest)  
\nonumber
\\
+ {} & \sum_{a = 1}^{5} F(a) \, G_{4} (\rest)
\nonumber
\\
+ {} & \sum_{a = 1}^{4} \sum_{b = a + 1}^{5} \!\!\! G_{2} (a,b) \, G_{3} (\rest)
\nonumber
\\
+ {} & G_{5} (1,2,3,4,5) ,
\label{cluster_F5}
\end{align}
with the periodic convention that ${ 6 \!=\! 1 }$, ${ 7 \!=\! 2 }$...
We note that these definitions ensure that all functions
are symmetric w.r.t.\ the interchange of any two coordinates.

To check the sanity of Eqs.~\eqref{cluster_F2}--\eqref{cluster_F5},
one can compute the ``norm''  of the correlation functions, ${G_{n}}$,
by integrating over all their variables.
Recalling that~\citep[see, e.g.\@, \S{A1} in][]{Fouvry+2020}
\begin{equation}
\!\! \int \!\! \rd 1 ... \rd n \, F_{n} (1,...,n) = \frac{N!}{(N \!-\! n)!} \, \mu^{n} ,
\label{norm_Fn}
\end{equation}
with the notation ${\rd 1 \!=\! \rd \bw_{1} }$,
we find~\citep{MMA}
\begin{subequations}
\begin{align}
{} & \!\! \int \!\! \rd 1 \, F(1) = N \mu ,
\\
{} & \!\! \int \!\! \rd 1 \rd 2 \, G_{2} (1,2) = - N \mu^{2} ,
\\
{} & \!\! \int \!\! \rd 1 \rd 2 \rd 3 \, G_{3} (1,2,3) = 2 N \mu^{3} ,
\\
{} & \!\! \int \!\! \rd 1 \rd 2 \rd 3 \rd 4 \, G_{4} (1,2,3,4) = - 6 N \mu^{4} ,
\\
{} & \!\! \int \!\! \rd 1 \rd 2 \rd 3 \rd 4 \rd 5 \, G_{5} (1,2,3,4,5) = 24 N \mu^{5} .
\end{align}
\label{norm_Gn}\end{subequations}
Since ${ \mu \!\sim\! 1/N }$,
we have the expected scalings w.r.t.\ $N$, namely
${ F \!\sim\! 1 }$ and ${ G_{n} \!\sim\! 1/N^{n-1} }$.

The next step of the calculation is to inject the decompositions
from Eqs.~\eqref{cluster_F2}--\eqref{cluster_F5}
into Eq.~\eqref{BBGKY_Fn} to obtain evolution equations
for ${ \p F / \p t }$ and ${ \p G_{n} / \p t }$. These cumbersome
manipulations are more easily performed using a computer
algebra system, as explicitly detailed in~\citep{MMA}.

To perform a truncation of the evolution equations
at order ${1/N^{3}}$, we follow the same approach as in~\citep{Fouvry2022}
and introduce the small parameter ${ \veps \!=\! 1/N }$.
More precisely, we perform the replacements
\begin{subequations}
\begin{align}
\mu {} & \to \mu \, \veps ,
\\
G_{2} {} & \to \veps G_{2}^{(1)} + \veps^{2} G_{2}^{(2)} + \veps^{3} G_{2}^{(3)} ,
\\
G_{3} {} & \to \veps^{2} G_{3}^{(2)} + \veps^{3} G_{3}^{(3)} ,
\\
G_{4} {} & \to \veps^{3} G_{4}^{(3)} ,
\\
G_{5} {} & \to \veps^{4} G_{5}^{(4)} .
\end{align}
\label{replacement_eps}\end{subequations}
Once these replacements performed,
we keep only terms up to order $\veps^3$,
and gather the terms order by order in $\veps$.
We are left with a series of evolution equations
for the various correlation functions~\citep{MMA}.

Assuming that ${ F (J,t) }$ and ${ \rd J }$
are all of size $1$ w.r.t.\ $N$ and $G$,
and introducing ${ U \!\propto\! G }$ as the typical scale
of the interaction potential, we are left with evolution equations
scaling like
\begin{subequations}
\begin{align}
\frac{\p F}{\p t} {} & \!=\! U G_{2}^{(1)} \!+\! U G_{2}^{(2)}  \!+\! U G_{2}^{(3)} ,
\\
\frac{\p G_{2}^{(1)}}{\p t} \!+\! \Omega G_{2}^{(1)} {} & \!=\! U G_{2}^{(1)} \!+\! \mu U ,
\\
\frac{\p G_{2}^{(2)}}{\p t} \!+\! \Omega G_{2}^{(2)} {} & \!=\! U G_{2}^{(2)} \!+\! \mu U G_{2}^{(1)} + U G_{3}^{(2)},
\\
\frac{\p G_{2}^{(3)}}{\p t} \!+\! \Omega G_{2}^{(3)} {} & \!=\! U G_{2}^{(3)} \!+\! \mu U G_{2}^{(2)} \!+\! U G_{3}^{(3)} ,
\\
\frac{\p G_{3}^{(2)}}{\p t} \!+\! \Omega G_{3}^{(2)} {} & \!=\! U G_{3}^{(2)} \!+\! \mu U G_{2}^{(1)} \!+\! U G_{2}^{(1)} G_{2}^{(1)} ,
\\
\frac{\p G_{3}^{(3)}}{\p t} \!+\! \Omega G_{3}^{(3)} {} & \!=\! U G_{3}^{(3)} \!+\! \mu U G_{2}^{(2)} \!+\! U G_{2}^{(1)} G_{2}^{(2)}
\nonumber
\\
{} & \; \!+\! \mu U G_{3}^{(2)} \!+\! U G_{4}^{(3)} ,
\\
\frac{\p G_{4}^{(3)}}{\p t} \!+\! \Omega G_{4}^{(3)} {} & \!=\! U G_{4}^{(3)} \!+\! \mu U G_{3}^{(2)} \!+\! \mu U G_{2}^{(1)} G_{2}^{(1)}
\nonumber
\\
{} & \; \!+\! U G_{2}^{(1)} G_{3}^{(2)} .
\end{align}
\label{hierarchy}\end{subequations}
Deriving a kinetic equation for ${ \p F / \p t }$
amounts to solving, in sequence,
this intricate hierarchy of coupled
partial integro-differential equations.

Placing ourselves in the limit ${ U \!\ll\! 1 }$,
i.e.\ neglecting collective effects,
and assuming that the four-body correlation function,
$G_{4}^{(3)}$, drives the dynamics,
we find
that the system's evolution is sourced by the sequence
of resolutions\footnote{Unfortunately,
the present argument is only heuristic,
as we can only guess that some other sequence of resolutions,
e.g.\@, ${ G_{2}^{(1)} \!\to\! G_{2}^{(2)} \!\to\! G_{2}^{(3)} \!\to\! \p F / \p t }$,
will lead to vanishing contributions by symmetry.
Though, the sequence from Eq.~\eqref{sequence_resolution}
appears as the natural ${ 1/N^{3} }$ generalisation
of the ${1/N^{2}}$ sequence, ${ G_{2}^{(1)} \!\to\! G_{3}^{(2)} \!\to\! G_{2}^{(2)} \!\to\! \p F / \p t }$, 
used to derive Eq.~\eqref{exp_Landau_1N2}~\citep{Fouvry2022}.}
\begin{equation}
G_{2}^{(1)} \!\to\! G_{3}^{(2)} \!\to\! G_{4}^{(3)} \!\to\! G_{3}^{(3)} \!\to\! G_{2}^{(3)} \!\to\! \p F / \p t .
\label{sequence_resolution}
\end{equation}
Following Eq.~\eqref{hierarchy}, we then successively
find the scalings~\citep{MMA}
\begin{subequations}
\begin{align}
G_{2}^{(1)} {} & = \mu U ,
\\
G_{3}^{(2)} {} & = \mu^{2} U^{2} ,
\\
G_{4}^{(3)} {} & = \mu^{3} U^{3} ,
\\
G_{3}^{(3)} {} & = \mu^{3} U^{4} ,
\\
G_{2}^{(3)} {} & = \mu^{3} U^{5} ,
\\
\p F / \p t {} & = \mu^{3} U^{6} ,
\end{align}
\label{scalings_hierarchy}\end{subequations}
with ${ U \!\propto\! G }$, the amplitude of the pairwise coupling.
As a conclusion, in the hot limit,
the relaxation driven
by four-body correlations,
i.e.\ by ${ G_{4}^{(3)} }$,
is associated with the (very) long relaxation time
${ \Trelax \!\propto\! N^{3} / G^{6} }$.

\section{Numerical simulations}
\label{app:NBody}

Our goal is to integrate a set of $N$ particles
governed by the Hamiltonian from Eq.~\eqref{def_Htot},
with the particular interaction potential
from Eq.~\eqref{def_U_B2}.

\subsection{Equations of motion}
\label{app:EOM}

Following Hamilton's equations,
the motion follows
\begin{subequations}
\begin{align}
\frac{\rd \theta_{i}}{\rd t} {} & = \Oext (J_{i}) + G \sum_{\mathclap{j \neq i}}^{N} \mu \, 2 \, (J_{i} \!-\! J_{j}) \, \mB_{2} [ \theta_{i} \!-\! \theta_{j}] ,
\label{EOM_theta}
\\
\frac{\rd J_{i}}{\rd t} {} & = - G \sum_{\mathclap{j \neq i}}^{N} \mu \, (J_{i} \!-\! J_{j})^{2} \, \mB_{2}^{\prime} [ \theta_{i} \!-\! \theta_{j} \big] ,
\label{EOM_J}
\end{align}
\label{EOM_thetaJ}\end{subequations}
with ${ \Oext \!=\! \rd \Uext / \rd J }$.
Such a system has two invariants
\begin{subequations}
\begin{align}
\Etot {} & = \sum_{i = 1}^{N} \mu \, \Uext (J_{i}) + \sum_{i < j}^{N} \mu^{2} \, U(\bw_{i} , \bw_{j}) ,
\label{def_Etot}
\\
\Jtot {} & = \frac{1}{\Mtot} \sum_{i = 1}^{N} \mu \, J_{i} .
\label{def_Jtot}
\end{align}
\label{def_Etot_Jtot}\end{subequations}

Owing to the system's ${2\pi}$-periodicity,
we always rewrap the angles ${ \theta_{i} }$
within the interval ${ [0,2\pi] }$.
In that case, the wrapping function from Eq.~\eqref{def_wrapping}
reads
\begin{equation}
w_{2 \pi} [\theta \!-\! \thetap] = | \theta \!-\! \thetap | =
\begin{cases}
\displaystyle \theta \!-\! \thetap {} & \mathrm{if} \quad \thetap < \theta ,
\\
\displaystyle \thetap \!-\! \theta {} & \mathrm{if} \quad \theta < \thetap ,
\end{cases}
\label{simple_w}
\end{equation}
and similarly for its gradient
\begin{equation}
\frac{\rd w_{2 \pi} [\theta \!-\! \thetap]}{\rd \theta} = \Sign [ \theta \!-\! \thetap] = 
\begin{cases}
\displaystyle 1 {} & \mathrm{if} \quad \thetap < \theta ,
\\
\displaystyle -1 {} & \mathrm{if} \quad \theta < \thetap .
\end{cases}
\label{simple_wp}
\end{equation}
Two important remarks can be made from Eqs.~\eqref{simple_w}
and~\eqref{simple_wp}.
First, provided that the respective order of ${ (\theta , \thetap) }$
is known, these two equations are separable w.r.t.\ both angles.
This will enable an accelerated evaluation of the forces, see Appendix~\ref{app:Multipole}.
Second, the pairwise force is discontinuous when two particles
cross in angle-space. As a consequence, traditional high-order Runge-Kutta
integration schemes can only be first-order accurate.
These discontinuities
are the main source of errors when integrating the system's dynamics,
see Appendix~\ref{app:TimeIntegration}.

\subsection{``Multipole'' acceleration}
\label{app:Multipole}

A naive implementation of Eq.~\eqref{EOM_thetaJ} 
requires ${ \mO (N^{2}) }$ evaluations.
Fortunately, Eqs.~\eqref{simple_w} and~\eqref{simple_wp} being ``quasi-separable'',
this evaluation can be accelerated using a multipole-like method.

Recalling that the angles $\theta_{i}$
are always rewrapped to the interval ${ [0 , 2 \pi] }$,
we first sort the particles according to $\theta_{i}$,
in ${ \mO (N \ln N) }$ operations.
Then, Eq.~\eqref{EOM_thetaJ}, once explicitly expanded,
involves polynomials of second-order
in ${ ( \theta_{i}, J_{j} ) }$.
Defining the upper/lower moments as
\begin{equation}
P_{k\kp} (\theta_{i}) \!=\! \sum_{\mathclap{\theta_{j} < \theta_{i}}} \mu \, \theta_{j}^{k} \, J_{j}^{\kp} ;
\;
Q_{k\kp} (\theta_{i}) \!=\! \sum_{\mathclap{\theta_{j} > \theta_{i}}} \mu \, \theta_{j}^{k} \,  J_{j}^{\kp} ,
\label{def_PQ}
\end{equation}
we can rewrite Eq.~\eqref{EOM_thetaJ} as
\begin{subequations}
\begin{align}
\frac{\rd \theta_{i}}{\rd t} {} & = \Oext (J_{i}) + \frac{\rd \theta_{i}}{\rd t} \bigg|_{P} + \frac{\rd \theta_{i}}{\rd t} \bigg|_{Q} ,
\label{expanded_EOM_theta}
\\
\frac{\rd J_{i}}{\rd t} {} & = \frac{\rd J_{i}}{\rd t} \bigg|_{P} + \frac{\rd J_{i}}{\rd t} \bigg|_{Q} .
\label{expanded_EOM_J}
\end{align}
\label{expanded_EOM}\end{subequations}
In these expressions, the upper/lower contributions read
\begin{subequations}
\begin{align}
\frac{1}{G} \frac{\rd \theta_{i}}{\rd t} \bigg|_{P} = {} & \big( \tfrac{1}{3} \!-\! \tfrac{1}{\pi} \theta_{i} \!+\! \tfrac{1}{2 \pi^{2}} \theta_{i}^{2} \big) \big( J_{i} P_{00} \!-\! P_{01} \big)
\label{dthetadt_P}
\\
+ {} & \tfrac{1}{\pi} \big( 1 \!-\! \tfrac{1}{\pi} \theta_{i} \big) \big( J_{i} P_{10} \!-\! P_{11}) \!+\! \tfrac{1}{2 \pi^{2}} \big( J_{i} P_{20} \!-\! P_{21} \big) ,
\nonumber
\\
\frac{1}{G} \frac{\rd \theta_{i}}{\rd t} \bigg|_{Q} = {} & \big( \tfrac{1}{3} \!+\! \tfrac{1}{\pi} \theta_{i} \!+\! \tfrac{1}{2 \pi^{2}} \theta_{i}^{2} \big) \big( J_{i} Q_{00} \!-\! Q_{01} \big) 
\label{dthetadt_Q}
\\
- {} & \tfrac{1}{\pi} \big( 1 \!+\! \tfrac{1}{\pi} \theta_{i} \big) \big( J_{i} Q_{10} \!-\! Q_{11}) \!+\! \tfrac{1}{2 \pi^{2}} \big( J_{i} Q_{20} \!-\! Q_{21} \big) ,
\nonumber
\end{align}
\label{dthetadt_PQ}\end{subequations}
as well as 
\begin{subequations}
\begin{align}
\frac{1}{G} \frac{\rd J_{i}}{\rd t} \bigg|_{P} = {} & \tfrac{1}{2 \pi} J_{i} \big( 1 \!-\! \tfrac{1}{\pi} \theta_{i} \big) \big( J_{i} P_{00} \!-\! 2 P_{01} \big) 
\label{dJdt_P}
\\
+ {} & \tfrac{1}{2 \pi} \big( 1 \!-\! \tfrac{1}{\pi} \theta_{i} \big) P_{02} \!+\! \tfrac{1}{2 \pi^{2}} \big( J_{i}^{2} P_{10} \!-\! 2 J_{i} P_{11} \!+\! P_{12} \big) ,
\nonumber
\\
\frac{1}{G} \frac{\rd J_{i}}{\rd t} \bigg|_{Q} = {} & - \tfrac{1}{2 \pi} J_{i} \big( 1 \!+\! \tfrac{1}{\pi} \theta_{i} \big) \big( J_{i} Q_{00} \!-\! 2 Q_{01} \big)
\label{dJdt_Q}
\\
- {} & \tfrac{1}{2 \pi} \big( 1 \!+\! \tfrac{1}{\pi} \theta_{i} \big) Q_{02} \!+\! \tfrac{1}{2 \pi^{2}} \big( J_{i}^{2} Q_{10} \!-\! 2 J_{i} Q_{11} \!+\! Q_{12} \big) .
\nonumber
\end{align}
\label{dJdt_PQ}\end{subequations}
To compute the rates of change, one scrolls
through the angles-sorted particles in increasing order
[resp.\ decreasing order]
while accumulating the various ${ \{P_{k\kp} \} }$
[resp.\ ${ \{ Q_{k\kp} \} }$].
This is performed in ${ \mO (N) }$ operations.

\subsection{Time integration}
\label{app:TimeIntegration}

Because the interaction potential from Eq.~\eqref{def_U_B2}
has discontinuous derivatives,
it is ill-advised to use high-order integration schemes.
In practice, we use the explicit midpoint rule
with a fixed timestep, $h$.
To integrate ${ \dot{y} \!=\! f(y) }$
with ${ y \!=\! \{ \theta_{i} , J_{i} \}_{i} }$,
we compute the transformation ${ y (t \!=\! n h) \! = \! y_{n} \to y_{n+1} }$ via
\begin{subequations}
\begin{align}
f_{1} {} & = f (y_{n}) ,
\\
y_{1} {} & = y_{n} + \half h f_{1} ,
\\
f_{2} {} & = f(y_{1}) ,
\\
y_{n+1} {} & = y_{n} + h f_{2} .
\end{align}
\label{midpoint}\end{subequations}
Because the force ${ f(y) }$ is discontinuous,
the method from Eq.~\eqref{midpoint} is only first-order.
This requires therefore the use of very small timesteps, $h$,
as we now detail.

\subsection{Integration errors}
\label{app:ErrorsIntegration}

We are interested in (very) long-term simulations
for different values of ${ (N,G) }$.
We must therefore pick carefully our integration parameters
to ensure an appropriate conservation of the systems' invariants.

We first consider the relative error in $\Jtot$,
as defined in Eq.~\eqref{def_Jtot}.
Up to round-off errors, our computation of the rates of change
in Appendix~\ref{app:Multipole} conserves $\Jtot$,
i.e.\ ${ \sum_{i=1}^{N} \mu \dot{J_{i}} \!=\! 0 }$.
The midpoint method from Eq.~\eqref{midpoint} conserves linear invariants~\citep[see, e.g.\@,][]{Hairer+2006}. Therefore, the relative error in $\Jtot$
will grow from the (biased) accumulation at every timestep
of a round-off error of order ${ \eps_{0} \!\simeq\! 10^{-16} }$.
The final relative error in $\Jtot$
is then expected to scale like
\begin{equation}
\eps_{\Jtot}^{\final} \propto \eps_{0} \, \Tmax / h ,
\label{final_eps_Jtot}
\end{equation}
with $\Tmax$ the total integration time.

Let us now consider the relative error in $\Etot$,
as defined in Eq.~\eqref{def_Etot}.
For small enough timesteps,
errors will mainly accumulate when particles
``collide'' in $\theta$-space,
i.e.\ when their angles cross
therefore experiencing the discontinuity from Eq.~\eqref{simple_wp}.

Let us assume that particles 1 and 2 cross.
At the time of crossing, the respective force
between the two particles changes abruptly
by a factor of order ${G/N}$ (see Eq.~\ref{simple_wp}).
Then, the force on the two particles remains incorrect
of this amount for a duration of order $h$.
During one such timestep,
the typical error introduced in the phase space position
of the two particles is then ${ \eps_{12} \!\propto G h / N }$.
Since we are in the dynamically hot limit,
i.e.\ ${ G \!\to\! 0 }$,
the main source
of error in Eq.~\eqref{def_Etot} stems from the total kinetic energy.
Hence, after one collision, the relative error
introduced in $\Etot$ is of order
${ \eps_{\Etot} \propto \eps_{12}/N }$.
There are about ${ \mO (N^{2}) }$ angle-collisions
per dynamical time (which we take to be unity).
Assuming that errors accumulate in quadrature,
we finally find that final relative error in $\Etot$
scales like
\begin{equation}
\eps_{\Etot} \propto G h \Tmax^{1/2} / N .
\label{final_eps_Etot}
\end{equation}

In practice, ensuring a good enough
conservation of $\Etot$ is the main constraint
to consider when setting up the simulations.

\subsection{Simulation setups}
\label{app:Setups}

For a given frequency profile,
we need to ensure that our final errors in $\Etot$
remain similar as one varies ${ (N,G) }$.
For all the simulations,
we always consider the same initial \DF\@,
namely
\begin{equation}
F (J) \propto \re^{- J^{4}} ,
\label{def_Finit}
\end{equation}
correctly normalised.
This \DF\ is picked because it is not the steady state
of any of the external potentials from Eq.~\eqref{freq_profile}.
We appropriately tune
the total integration time, $\Tmax$,
and the integration timestep, $h$, as follows.

\textbf{Integration times}. In order to observe relaxation,
we take ${ \Tmax \!\propto\! \Trelax }$
(Eq.~\ref{Trelax_123}).
We therefore choose
\begin{subequations}
\begin{align}
{} & \quad \text{(1):} \;\;\; \Tmax \propto N/G^{2} ;
\label{Tmax_1}
\\
{} & \quad \text{(2):} \;\;\; \Tmax \propto N^{2}/G^{4} ;
\label{Tmax_2}
\\
{} & \quad \text{(3):} \;\;\; \Tmax \propto N^{2}/G^{6} .
\label{Tmax_3}
\end{align}
\label{Tmax_123}\end{subequations}

\textbf{Integration timesteps}. We fix $h$
so that the final relative error in $\Etot$
(Eq.~\ref{final_eps_Etot})
is independent of ${ (N,G) }$.
In practice, we find
\begin{subequations}
\begin{align}
{} & \quad \text{(1):} \;\;\; h \propto N^{1/2} ;
\label{h_1}
\\
{} & \quad \text{(2):} \;\;\; h \propto G ;
\label{h_2}
\\
{} & \quad \text{(3):} \;\;\; h \propto G^{2} .
\label{h_3}
\end{align}
\label{h_123}\end{subequations}

\textbf{Computational cost}. The difficulty of integrating
for one timestep scales like
${ \mO (N \ln N) }$ (Appendix~\ref{app:Multipole}). 
Hence, the difficulty of performing one full integration
scales like ${ \mO (N \Tmax \ln N / h) }$.
For the different frequency profiles,
we find
\begin{subequations}
\begin{align}
{} & \quad \text{(1):} \;\;\; \Diff \propto N^{3/2} \ln N / G^{2} ;
\label{Diff_1}
\\
{} & \quad \text{(2):} \;\;\; \Diff \propto N^{3} \ln N / G^{5} ;
\label{Diff_2}
\\
{} & \quad \text{(3):} \;\;\; \Diff \propto N^{3} \ln N / G^{8} .
\label{Diff_3}
\end{align}
\label{Diff_123}\end{subequations}
As expected, the integrations for the profile (3)
are the most challenging ones.

Following these guidelines,
to obtain Fig.~\ref{fig:Scaling},
we used the integration parameters
from Table~\ref{table:B2}.
\begin{table}[htbp!]
\begin{center}
\begin{tabular}{| p{0.15\textwidth} || p{0.3\textwidth} |}
\hline
All profiles &
 \\
\hline
${ \Gmax }$ &
${ \Gcrit/3 }$
\\
\hline
$G$ &
${ 2^{-n/4} \!\times\! \Gmax \quad (0 \!\leq\! n \!\leq\! 4) }$
\\
\hline
$N$ &
${ \lfloor 2^{n/4} \!\times\! \Nmin \rceil \quad  (0 \!\leq\! n \!\leq\! 4) }$
\\
\hline
\end{tabular}
\\[5pt]
\begin{tabular}{| p{0.15\textwidth} || p{0.3\textwidth} |}
\hline
Profile &
(1)
\\
\hline
$\Gcrit ; \Nmin $ &
1.85; 1024
\\
\hline
Timestep $h$ &
${ (N / \Nmin)^{1/2} / 50000}$
\\
\hline
Total time $\Tmax$ &
${ 10^{2} \!\times\! (N / \Nmin) \, (G / \Gmax)^{-2} }$
\\
\hline
\end{tabular}
\\[5pt]
\begin{tabular}{| p{0.15\textwidth} || p{0.3\textwidth} |}
\hline
Profile &
(2)
\\
\hline
${ \Gcrit; \Nmin }$ &
8.95; 32
\\
\hline
Timestep $h$ &
${ (G / \Gmax) / 3500}$
\\
\hline
Total time $\Tmax$ &
${ 2.3 \!\times\! 10^{5} \!\times\! (N / \Nmin)^{2} \, (G / \Gmax)^{-4} }$
\\
\hline
\end{tabular}
\\[5pt]
\begin{tabular}{| p{0.15\textwidth} || p{0.3\textwidth} |}
\hline
Profile &
(3)
\\
\hline
${ \Gcrit; \Nmin }$ &
9.87; 32
\\
\hline
Timestep $h$ &
${ (G / \Gmax)^{2} / 1500}$
\\
\hline
Total time $\Tmax$ &
${ 9.3 \!\times\! 10^{4} \!\times\! (N / \Nmin)^{2} \, (G / \Gmax)^{-6} }$
\\
\hline
\end{tabular}
\caption{Integration parameters for the profiles
from Eq.~\eqref{freq_profile}.
\label{table:B2}}
\end{center}
\end{table}
For every frequency profile
and every value of ${ (N,G) }$,
we perform a total of ${ \Nreal \!=\! 10^{3} }$ realisations.
These are time-consuming simulations.
Indeed, on a 128-core CPU,
running 50 independent realisations
for all 25 values of ${ (N,G) }$
required ${ \!\sim\!3 \, \hr}$
[resp.\ ${ \!\sim\!140 \, \hr }$; ${ \!\sim\!150 \, \hr }$]
for the profile (1) [resp.\ (2); (3)].

\begin{figure}[htbp!]
    \begin{center}
\includegraphics[width=0.45\textwidth]{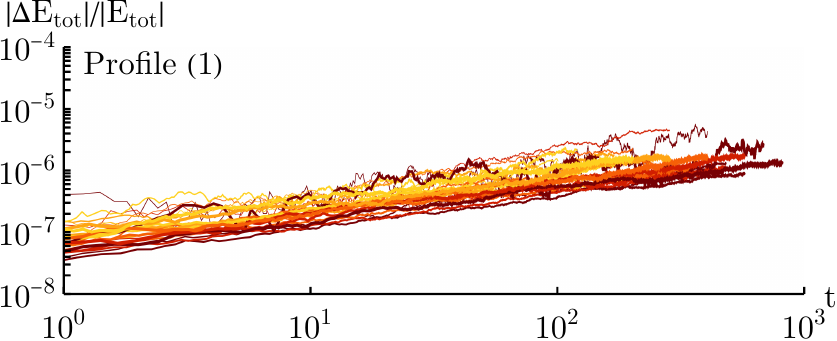}
\includegraphics[width=0.45\textwidth]{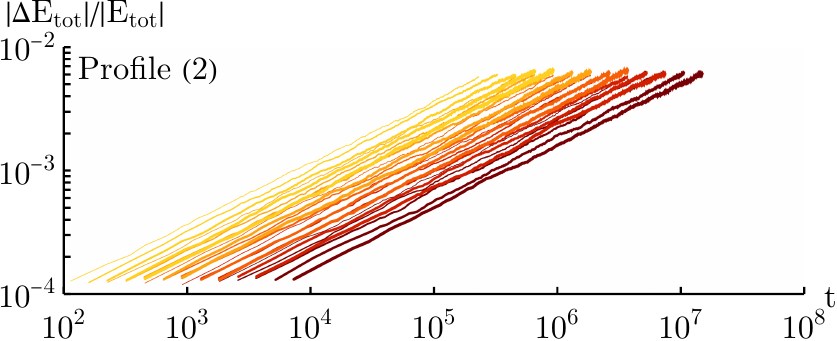}
\includegraphics[width=0.45\textwidth]{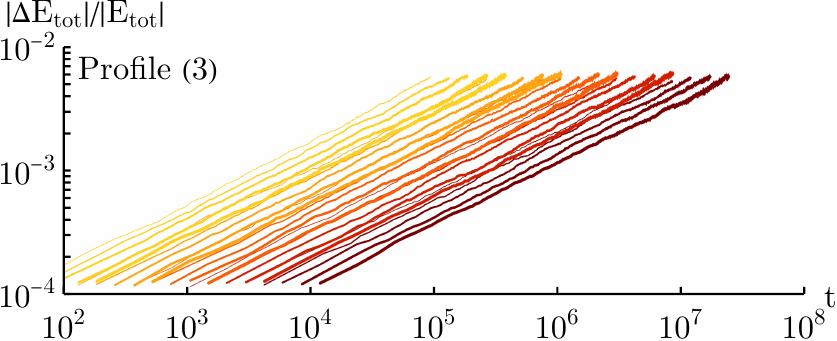}
    \caption{Time evolution of the relative error in $\Etot$
for the frequency profiles from Eq.~\eqref{freq_profile}
(top to bottom panels), as one increases $N$ (thin to thick)
and increases $G$ (red to yellow).
    \label{fig:IntegrationErrorEtot}}
    \end{center}
\end{figure}

\begin{figure}[htbp!]
    \begin{center}
\includegraphics[width=0.45\textwidth]{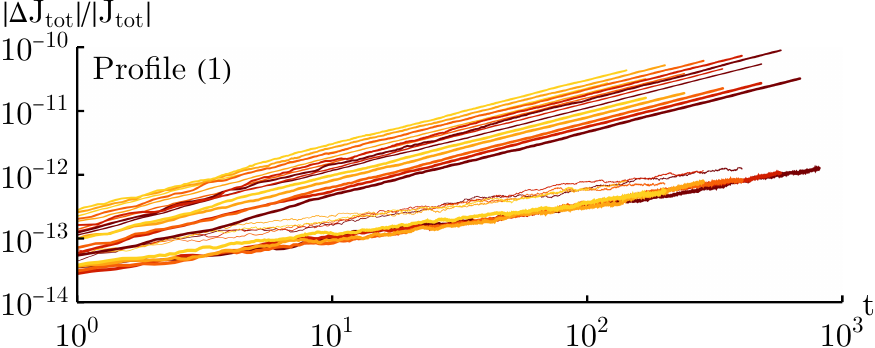}
\includegraphics[width=0.45\textwidth]{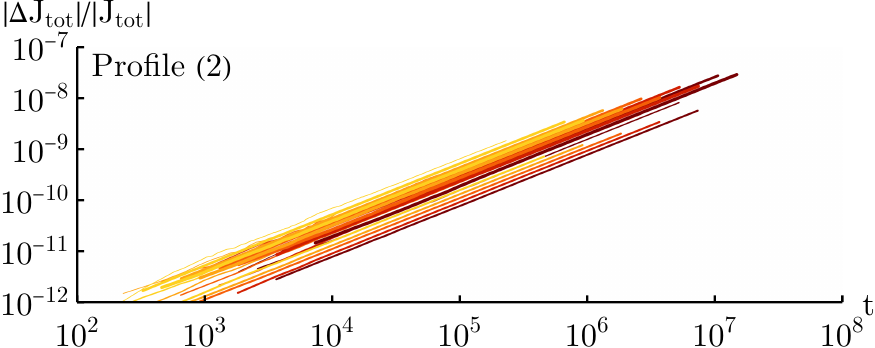}
\includegraphics[width=0.45\textwidth]{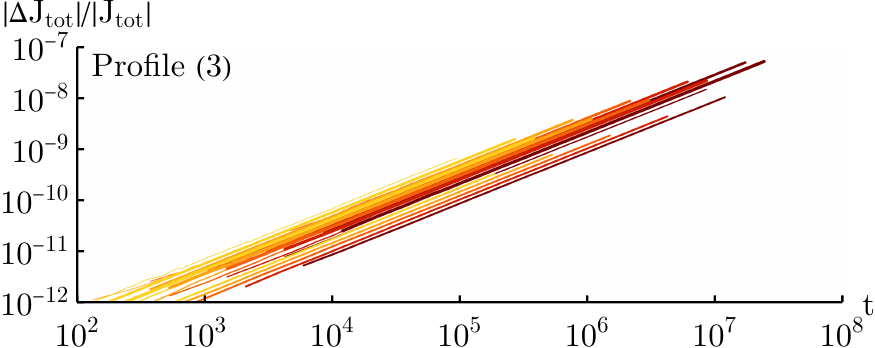}
    \caption{Same as Fig.~\ref{fig:IntegrationErrorEtot}
but for $\Jtot$.
    \label{fig:IntegrationErrorJtot}}
    \end{center}
\end{figure}

In Fig.~\ref{fig:IntegrationErrorEtot}
[resp.\ Fig.~\ref{fig:IntegrationErrorJtot}],
we illustrate the time-dependence of the relative errors
in $\Etot$ [resp.\ $\Jtot$].
As expected, simulations with different values
of ${ (N,G) }$ show similar relative errors in $\Etot$
at the end of their respective integrations.

\subsection{Measuring relaxation}
\label{app:Measurements}

We track relaxation via centred moments in action.
Taking inspiration from~\cite{Joyce+2010,Rocha+2014},
we consider
\begin{equation}
m_{4} (t) = \frac{1}{\Mtot} \sum_{i = 1}^{N} \mu \, \big( J_{i} \!-\! \Jtot \big)^{4} ,
\label{def_m4}
\end{equation}
where, up to integration errors,
$\Jtot$ is conserved in every realisation (Eq.~\ref{def_Jtot}).

\begin{figure}[htbp!]
    \begin{center}
\includegraphics[width=0.45\textwidth]{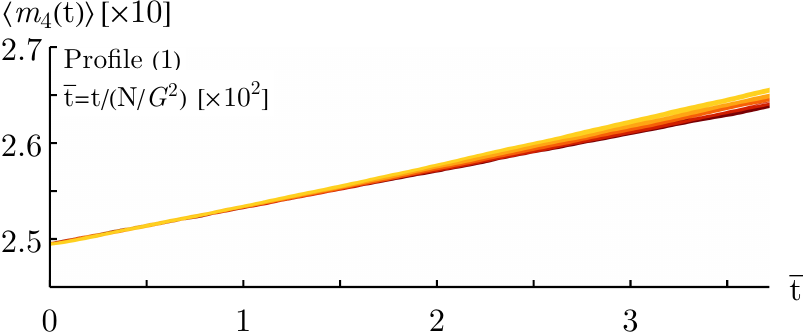}
\includegraphics[width=0.45\textwidth]{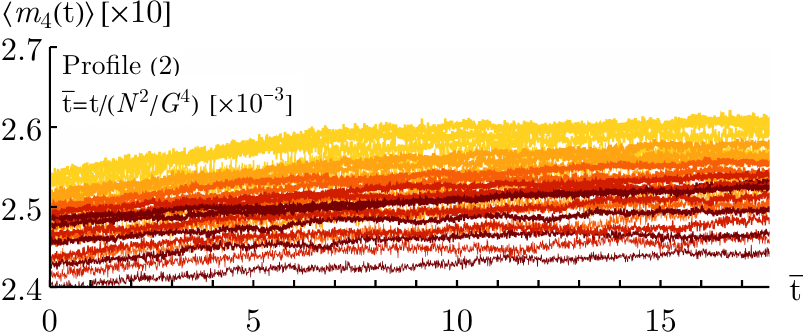}
\includegraphics[width=0.45\textwidth]{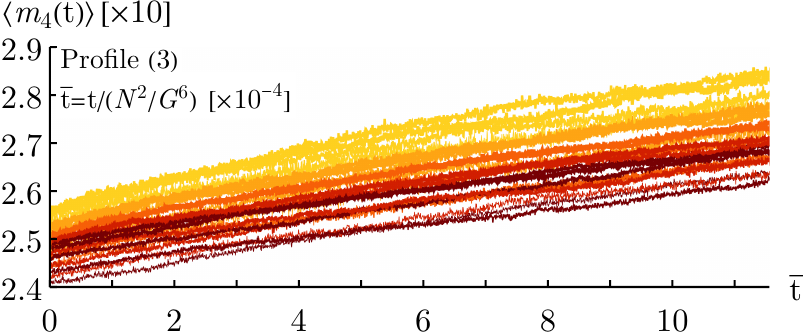}
    \caption{Time evolution of the ensemble-averaged
moment, ${ \langle m_{4} (t) \rangle }$,
for the different frequency profiles from Eq.~\eqref{freq_profile}
(top to bottom panels), as one increases $N$ (thin to thick)
and increases $G$ (red to yellow).
The time axis has been rescaled following
the expected relaxation time from Eq.~\eqref{Trelax_123}.
    \label{fig:m4}}
    \end{center}
\end{figure}

For a given frequency profile
and a given value of ${ (N,G) }$,
we have at our disposal a large set of realisations.
Performing an ensemble-average over realisations,
we are left with time series of the form
${ t \!\mapsto\! \langle m_{4} (t) \rangle }$,
as illustrated in Fig.~\ref{fig:m4}.
During the first few dynamical times,
the systems undergo a (slight) initial violent relaxation~\citep{LyndenBell1967}.
To prevent it from polluting our measurements,
we only consider the signals
for ${ t \!\geq\! 5 }$.

\begin{figure}[htbp!]
    \begin{center}
\includegraphics[width=0.45\textwidth]{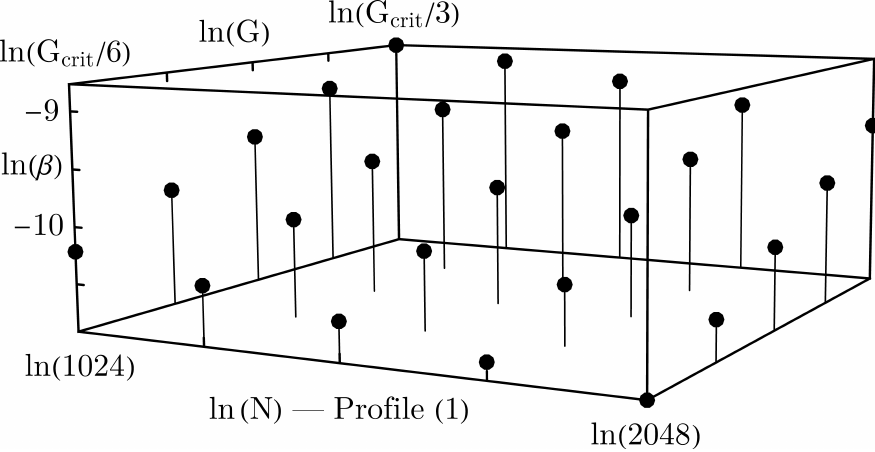}
\includegraphics[width=0.45\textwidth]{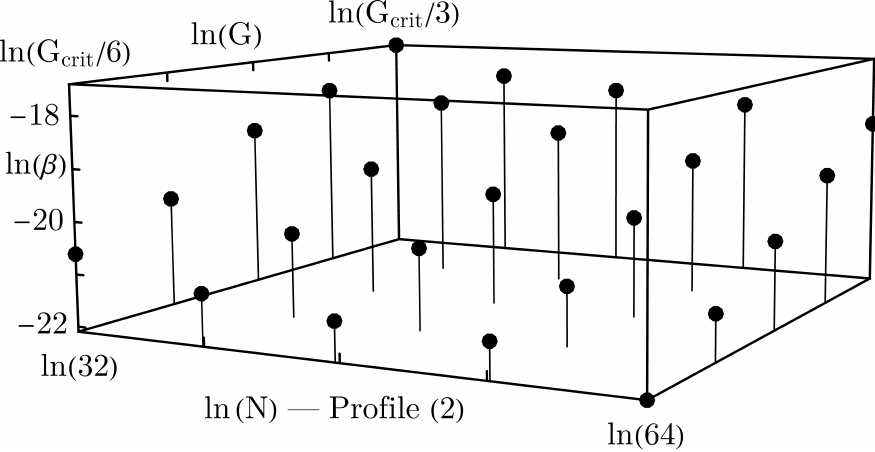}
\includegraphics[width=0.45\textwidth]{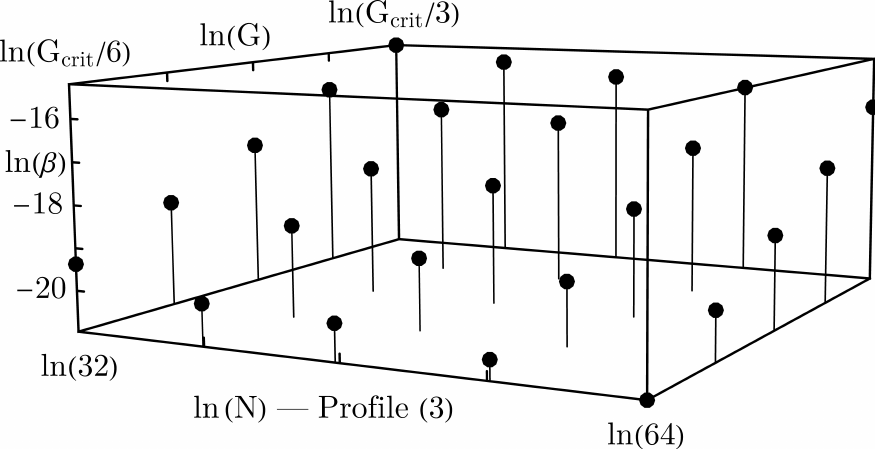}
    \caption{Dependence of the slope ${ \ln \beta (N,G) }$
for the different frequency profiles from Eq.~\eqref{freq_profile}
(top to bottom panels), as a function of ${(N,G)}$.
    \label{fig:beta}}
    \end{center}
\end{figure}

The crux of the measurement is then as follows.
For a given time series, we perform a linear fit
of the form
\begin{equation}
\langle m_{4} (t) \rangle \simeq \beta \, t + \cst,
\label{fit_m4}
\end{equation}
so that the slope ${ \beta \!=\! \beta (N,G) }$
is expected to be proportional to ${ 1/\Trelax }$.
In Fig.~\ref{fig:beta}, we illustrate the dependence
of ${\beta}$ as one varies ${(N,G)}$.
Then, having at our disposal a set of ${ \beta (N,G) }$,
we perform a linear fit of the form
\begin{equation}
\ln \beta (N,G) \simeq - \gamma_{N} \, \ln N + \gamma_{G} \, \ln G + \cst
\label{fit_beta}
\end{equation}
The coefficients ${ (\gamma_{N} , \gamma_{G}) }$
are the ones from Eq.~\eqref{ansatz_Trelax}.

In practice, it is important to estimate the uncertainties
of this measurement.
This is performed using bootstraps.
A given bootstrap proceeds as follows.
For every value of ${ (N,G) }$,
we perform the ensemble average of ${ m_{4} (t) }$
over a sample of ${ \Nreal }$ realisations
drawn, with repetitions, from the $\Nreal$ realisations available.
Using the associated ${ \langle m_{4} (t) \rangle }$,
we perform the linear fit from Eq.~\eqref{fit_m4}
from which we also estimate the variance of ${\beta}$.
For every value of ${ (N,G) }$, we draw a value
of ${ \beta (N,G) }$ according to the associated Gaussian distribution.
Finally, from this sample of ${ \beta (N,G) }$,
we perform the linear fit from Eq.~\eqref{fit_beta}.
Relying on the variance-covariance of ${ (\gamma_{N} , \gamma_{G}) }$,
we can draw one value for ${ (\gamma_{N}, \gamma_{G}) }$,
following the associated Gaussian distribution.
This constitutes one bootstrap measurement.
In practice, to obtain Fig.~\ref{fig:Scaling},
we performed a total of 1280 bootstrap measurements.

Once these samples of ${ \bgamma \!=\! (\gamma_{N} , \gamma_{G}) }$ available,
we estimate their associated PDF, ${ P (\bgamma) }$,
using \texttt{Mathematica}'s default \texttt{SmoothKernelDistribution}.
Finally, a contour labelled ${ x \% }$ in Fig.~\ref{fig:Scaling}
corresponds to the level line ${ P (\bgamma) \!=\! p }$,
with $p$ set by ${ Q(p) / Q(p\!=\!0) \!=\! x \% }$
and ${ Q (p) \!=\! \!\int_{P(\bgamma) \geq p} \rd \bgamma P (\bgamma) }$.

\subsection{Bias in $\gamma_{G}$}
\label{app:Bias}

In Fig.~\ref{fig:Scaling}, the kinetic prediction for $\gamma_{G}$
corresponds to the limit ${ G \!\to\! 0 }$,
while the measurements are performed
for finite values of $G$.
This leads to a biased overestimation of $\gamma_{G}$,
once again associated with leaks from collective effects.
In this section, we briefly estimate the maximum extent
of that pollution.

For a fixed value of $N$
the scaling of the ${1/N}$ \BL\ equation
w.r.t.\ $G$ is, roughly,
\begin{equation}
\frac{\p F}{\p t} \propto \bigg( \frac{G}{1 \!-\! G/\Gcrit} \bigg)^{2} .
\label{scaling_G_1/N}
\end{equation}
Here, we followed Eq.~\eqref{exp_Ud},
and wrote the dressed coupling coefficient as
${ \Ud\!\propto\! U / |\bE| \!\propto\! G / (1\!-\! G / \Gcrit)}$,
and subsequently abruptly neglected the frequency dependence
of the dielectric matrix, ${ \bE (\omega) }$.

Similarly, within the same limits
and following Eq.~\eqref{secondorder_Landau_eq},
the ${ 1/N^{2} }$ \BL\ is expected to scale w.r.t.\ $G$ roughly like
\begin{equation}
\frac{\p F}{\p t} \propto \bigg( \frac{G}{1 \!-\! G/\Gcrit} \bigg)^{4} .
\label{scaling_G_1/N2}
\end{equation}

When measuring numerically relaxation rates,
we limited ourselves to the values
${ G/\Gcrit \!=\! \{ 2^{-n/4} / 3 \} }$ with ${ 0 \!\leq\! n \!\leq\! 4 }$,
as detailed in Table~\ref{table:B2}.
In order to estimate the maximum bias in $\gamma_{G}$
associated with this particular choice,
we can compute ${ \p F / \p t }$ 
as given by Eqs.~\eqref{scaling_G_1/N} and~\eqref{scaling_G_1/N2},
and perform the linear fit 
${ \ln (\p F / \p t) \!\simeq\! \tgamma_{G} \ln G \!+\! \cst}$
In that case, ${ \tgamma_{G} }$ is then an estimate
of the maximum value of $\gamma_{G}$
that could stem from our use of finite values of $G$.
In practice, for the ${ 1/N }$ [resp.\ ${ 1/N^{2} }$] dynamics from Eq.~\eqref{scaling_G_1/N} [resp.\ Eq.~\eqref{scaling_G_1/N2}],
we find ${ \tgamma_{G} \!\simeq\! 2.64 }$ [resp.\ ${ \tgamma_{G} \!\simeq\! 5.28 }$].
Fortunately, these values of $\tgamma_{G}$
are larger than the mean values obtained in Fig.~\ref{fig:Scaling}.
This strengthens our confidence in the sanity
of the numerical measurements.

\end{document}